\documentclass[12pt]{iopart}

\eqnobysec

\usepackage{graphicx}

\font\amsb=msbm10
\def\hbar{\mbox{\amsb\char'175}}

\newcommand{\be}{\begin{equation}}
\newcommand{\ee}{\end{equation}}

\newcommand{\vecp}{{\mathbf p}}

\newcommand{\vecq}{{\mathbf q}}

\newcommand{\x}{{\mathbf x}}

\newcommand{\X}{{\mathbf X}}

\newcommand{\y}{{\mathbf y}}

\newcommand{\vl}{{\mathbf l}}

\newcommand{\J}{{\mathbf J}}

\newcommand{\G}{{\mathbf G}}

\newcommand{\C}{{\mathbf C}}

\newcommand{\opA}{{\widehat{A}}}

\newcommand{\opB}{{\widehat{B}}}

\newcommand{\opR}{{\widehat{R}}}

\newcommand{\opT}{{\widehat{T}}}

\newcommand{\oprho}{{\widehat{\rho}}}

\newcommand{\mH}{{\mathbf H}}

\newcommand{\opH}{\widehat{H}}

\newcommand{\opU}{\widehat{U}}

\newcommand{\oq}{\widehat{q}}

\newcommand{\op}{\widehat{p}}

\newcommand{\opx}{\widehat{\mathbf x}}

\newcommand{\M}{{\mathbf M}}

\newcommand{\der}{\partial}

\newcommand{\vct}[1]{\ensuremath\mbox{\boldmath$ #1 $}}

\newcommand{\Vxi}{\vct \xi}

\newcommand{\Vl}{\vct l}

\begin{document}

\title{Semiclassical Evolution of Dissipative Markovian Systems}

\author {A. M. Ozorio de Almeida\footnote{Corresponding author}}
\address{Centro Brasileiro de Pesquisas Fisicas; \\
Rua Xavier Sigaud 150, 22290-180, Rio de Janeiro, RJ,
Brazil.\\ozorio@cbpf.br}

\author{P. de M. Rios}
\address{Departamento de Matem\'atica, ICMC, Universidade de S\~ao
Paulo;\\Cx Postal 668, 13560-970, S\~ao Carlos, SP,
Brazil.\\prios@icmc.usp.br}

\author{O. Brodier}
\address{Laboratoire de Math\'ematiques et Physique Th\'eorique,
\\
Universit\'e des Sciences et Techniques Universit\'e de Tours
Parc de Grandmont; \\
37200, Tours, France.\\brodier@lmpt.univ-tours.fr}

\begin{abstract}

A semiclassical approximation for
an evolving density operator, driven by a ``closed'' hamiltonian
operator and ``open'' markovian Lindblad operators, is obtained. 
The theory is based on the chord function,
i.e. the Fourier transform of the Wigner function.
It reduces to an exact solution of the Lindblad master equation if
the hamiltonian operator is a quadratic function
and the Lindblad operators are linear functions of positions and momenta.
 
Initially, the semiclassical formulae for the case 
of hermitian Lindblad operators are reinterpreted in terms of
a (real) double phase space, generated by an appropriate 
classical double Hamiltonian. 
An extra ``open'' term is added to the double Hamiltonian
by the non-hermitian part of the Lindblad operators
in the general case of dissipative markovian evolution. 
The particular case of generic hamiltonian operators, 
but linear dissipative Lindblad operators, is studied in more detail. 
A Liouville-type equivariance still holds 
for the corresponding classical evolution in double
phase, but the centre subspace, which supports the
Wigner function, is compressed, along with expansion of its
conjugate subspace, which supports the chord function.

Decoherence narrows the relevant region of double phase space 
to the neighborhood of a caustic for both the Wigner function 
and the chord function. This difficulty is avoided 
by a propagator in a mixed representation, so that
a further ``small-chord'' approximation leads to a simple
generalization of the quadratic theory for evolving Wigner
functions.

\end{abstract}

\maketitle

\section{Introduction}

The Lindblad master equation describes the general evolution for
markovian open systems under the weakest possible constraints
\cite{Lindblad} (see also e.g. \cite{Giulini, Percival}). Given
the internal Hamiltonian, $\hat{H}$, and the Lindblad operators,
$\hat{L_k}$, which account for the action of the random
environment, the evolution of the density operator may be reduced
to the canonical form,
\begin{equation}
\frac{\partial \hat{\rho}}{\partial t} = -\frac{i}{\hbar}[\hat{H},
\hat{\rho}] + \frac{1}{\hbar}\sum_k
(\hat{L_k}\hat{\rho}\hat{L_k}^{\dag} -
\frac{1}{2}\hat{L_k}^{\dag}\hat{L_k}\hat{\rho} - \frac{1}{2}
\hat{\rho}\hat{L_k}^{\dag}\hat{L_k} ), \label{Lindblad}
\end{equation}
so that, in the absence of the environment ($\hat{L_k}=0$), the
motion is governed by the Liouville-Von Neumann equation
appropriate for unitary evolution.

A typical example is based on the Jaynes-Cummings model, which
describes the interaction of a two-level atom with a single mode
of the optical field  in a cavity. The statistically independent
arrival of atoms leads to the {\it damped harmonic oscillator
equation} for the photon field,
\begin{eqnarray}
\frac{\partial \hat{\rho}}{\partial t} =
-\frac{i}{\hbar}[\hat{a}^{\dag}\hat{a},\hat{\rho}] &&+
\frac{A}{\hbar}(\nu+1) (\hat{a}\hat{\rho}\hat{a}^{\dag} -
\frac{1}{2}\hat{a}^{\dag}\hat{a}\hat{\rho} -
\frac{1}{2} \hat{\rho}\hat{a}^{\dag}\hat{a} ) \nonumber \\
&&+ \frac{A}{\hbar}\nu (\hat{a}^{\dag}\hat{\rho}\hat{a} -
\frac{1}{2}\hat{a}\hat{a}^{\dag}\hat{\rho} - \frac{1}{2}
\hat{\rho}\hat{a}\hat{a}^{\dag} ), \label{opt}
\end{eqnarray}
where we identify the pair of Lindblad operators as proportional
to the anihilation operator $\hat{a}=(\oq +i \op)/\sqrt{2}$ and
the creation operator $\hat{a}^{\dag}=(\oq -i \op)/\sqrt{2}$ for
photons in the field mode (see e.g. \cite{Schleich, Englert} and
references therein). Further examples, e.g. laser models and heavy
ions conditions, are reviewed in \cite{Isar94}.

It should be mentioned
that the Lindblad master equation is open to several criticisms.
It has been explicitly shown, in the case of 
the dampened harmonic oscillator, that there is no universal master equation
describing exactly the evolution of the reduced system, 
i.e. one that is fully independent of the initial state \cite{Karr97}.
It may further be argued that other choices for the master equation 
can be more appropriate \cite{Whitney2008}. Nonetheless, the theory of
semigroups is the most obvious generalization of the unitary evolution of
isolated systems. Beyond its proven usefulness, 
this singles out the particular Markovian structure 
as a fundamental subject for investigation. 

The case where the Lindblad operators are all self-adjoint has
deserved special attention. It is known that the corresponding
Lindblad equation describes decoherence, or dephasing, as well as
diffusion, but no dissipation \cite{Percival}. Since the latter is
usually a much slower process, it is often useful to simplify the
evolution by considering only the self-adjoint part of the
$\hat{L_j}$'s when studying the decoherence process (as in the
semiclassical theory proposed in \cite{OA03}). However, most
physical processes for an open system such as (\ref{opt}) are
dissipative. It is therefore desirable to develop a semiclassical
theory for the evolution of the density operator that combines the
description of both the initial decoherence process and the more
classical development of diffusion and dissipation.

In this paper, we develop a formalism for treating the
semiclassical limit of evolving density operators subject
to equation (\ref{Lindblad}), including the cases where the Lindblad
operators are not self-adjoint. By semiclassical, we mean generalized 
WKB expansions (see e.g. \cite{Maslov}), as opposed to simple power expansions in
$\hbar$. The present theory expands on our
phase-space treatment for the semiclassical evolution of
closed systems \cite{RiosOA} and of non-dissipative open markovian
systems \cite{OA03}. To this purpose, we will adapt the theory
developed in these papers (particularly \cite{OA03}) in two
respects.

Recalling that $\bf{R}^{2N}$ stands for a $(2N)$-dimensional phase
space, which is a symplectic vector space, $\{\x=(\vecp,
\vecq)\}$, first we switch from the Weyl representation, where
$\hat{\rho}$ is represented by the Wigner function $W(\x)$, to its
Fourier transformed representation, the chord representation,
where $\hat{\rho}$ is represented by the chord function,
$\chi(\Vxi)$, also known as the quantum characteristic function,
given by
\begin{equation}
\label{Fourier1} \chi(\Vxi) = \frac{1}{(2\pi\hbar)^N}\int d\x \;
W(\x)\; \exp{\{\frac{i}{\hbar}(\x\wedge \Vxi)\}} \ ,
\end{equation}
where we have used the skew product, \begin{equation} \x\wedge \x'=\sum_{n=1}^N
(p_n q'_n - q_n p'_n)= \J\>\x \cdot \x', \label{squew} \end{equation} which
also defines the skew symplectic matrix $\J$. The {\it chord},
$\Vxi = (\xi_p,\xi_q)$, is the Fourier conjugate variable of the
{\it centre} $\x$ and stands for a tangent vector in phase space,
as in the scheme for a Legendre transform. In contrast to the
Wigner function, the chord function is not necessarily real, but
its semiclassical expression is often similar to that of the
Wigner function, as discussed in \cite{AlmVal04, Zambrano}.

In all cases where the Hamiltonian operator is at most
quadratic in the momentum and position operators, $\opx=(\hat{p},
\hat{q})$, and the Lindblad operators are linear in $\opx$, as in
example (\ref{opt}), the Lindblad equation reduces to a
Fokker-Planck equation in the chord representation, which can be
solved exactly for any initial state \cite{BroAlm04}. Various
instances of this result have been previously reported, e.g.   
\cite{Agarwal71,DodMan85,Sand87}. Keeping to linear (but not
self-adjoint) Lindblad operators, we now obtain an appropriate
semiclassical generalization to the evolution of the chord
function, $\chi(\Vxi,t)$, for generic Hamiltonian operators, given
an initial pure state, $\chi(\Vxi,0)$. This is similar to the
theory for the evolution of the Wigner function in \cite{OA03}, in
which the Lindblad operators were assumed to be self-adjoint (no
dissipation). But the present treatment has the immediate
advantage of being exact in the quadratic case.

In fact, the inverse Fourier transform of the semiclassical
evolution for the chord function, evaluated within the stationary
phase approximation, produces the same semiclassical evolution for the
Wigner function as was presented in \cite{OA03}. However, this can
now be seen to be a poorer approximation than the
semiclassical chord function presented here. 
Indeed, the theory in \cite{OA03} does not
describe diffusion, which progressively coarse-grains the
Wigner function. This is clear from the general analysis of the quadratic
case \cite{BroAlm04}, or \cite{Isar94,Sandu87} in the case of
initial coherent states.

The second modification to the WKB semiclassical theory, which is
required for treating markovian dissipation, is more profound: 
We find that it is necessary to work in {\it double phase space},
$(\x,\Vxi)\in\bf{R}^{2N}\times\bf{R}^{2N}$. This is a natural
setting for the corresponding description of the semiclassical
evolution of the density operator \cite{Littlejohn95, emaranha},
or indeed, for the representation of general operators acting on
the Hilbert space of quantum states, considered as superpositions
of $|ket\rangle \langle bra|$ elements. Just as
an evolving quantum state, $|\psi\rangle$, corresponds to an
evolving submanifold in simple phase space, $\x\in\bf{R}^{2N}$,
the unitary evolution of a pure state density operator in a closed system,
that is, a projector, $|\psi\rangle \langle\psi|$, corresponds to
the evolution of a submanifold in double phase space. (In both
cases, the respective submanifold satisfies an appropriate
lagrangian property, to be specified). We thus obtain a formal
generalization of the WKB framework, where an approximate oscillating
solution of the Schr\"odinger equation is built from a classically
evolving lagrangian submanifold \cite{Maslov, GS, Gutzwiller}.

However, the restriction to a closed system 
(and hence unitary quantum evolution) imposes a severe limitation 
on the allowed form of the corresponding classical double 
Hamiltonian \cite{OsKon}.
The crucial point here is that an additional term in the double Hamiltonian
arises naturally, as a consequence of 
the semiclassical approximation for the open terms in the master equation.
This new term, which is responsible for dissipation,  
depends exclusively on the Lindblad operators and cancels
in the special case where these are self-adjoint. 
Moreover, the resulting description of the full semiclassical evolution
again coincides with the exact solution of the master equation in
the quadratic case.

Our use of classical double phase is limited to the semiclassical approximation.
No attempt has here been made to define a generalized quantum mechanics
that would correspond to classical double phase space. 
This could lead to a fully quantum path integral for markovian systems,
as an alternative to the one developed by Strunz \cite{Strunz}.
His approach relies on the position representation, with
the obervables defined in the Weyl representation. Several of
the ingredients in our theory already appear in Strunz's path integral,
though its semiclassical limit is expressed in terms of complex orbits,
whereas we deal only with real phase space propagation. 

This paper is divided in three parts.
In the initial sections 2-5, we review basic material and
reformulate the semiclassical theory for closed evolution
\cite{RiosOA} and nondissipative open markovian evolution
\cite{OA03}, within the chord representation, so as to perfectly
fit the exact quadratic results in \cite{BroAlm04}.

In the second part, the ingredients in the basic result, equation
(\ref{chit}), are reinterpreted within the double phase space
scenario. This leads to the identification of the dissipative
hamiltonian, in section 6, and the consequent semiclassical
treatment of dissipative markovian dynamics, in section 7 (for
linear Lindblad operators).

It turns out that the {\it classical region} of double phase
space, to which decoherence drives the evolution, 
projects singularly as a caustic onto the subspaces where
either the Wigner function, or the chord function are defined. 
For this reason, in the final part of this paper, section 8, our
semiclassical theory is adapted to the evolving {\it centre-chord
propagator} \cite{AlmBro06}, which takes an initial density
operator, expressed as a Wigner function, into a final chord
function, thus avoiding caustics for a finite time. 
This leads to a {\it small chord approximation} for the evolution of the
Wigner function itself.

\section{Review of the semiclassical theory for density operators}

The chord representation of an operator $\hat{A}$ on the Hilbert
space ${L}^2(\bf{R}^N)$ is defined via the decomposition of
$\hat{A}$ as a linear (continuous) superposition of {\it
translation operators},
\begin{equation}
\hat{T}_{\Vxi} = \exp{\{\frac{i}{\hbar}(\Vxi\wedge \hat{\x})\}} ,
\end{equation}
also known as displacement operators. 
Each of these corresponds classically to a uniform translation of
phase space points $\x_0 \in \bf{R}^{2N}$ by the vector $\Vxi \in
\bf{R}^{2N}$, that is: $\x_0 \mapsto \x_0 + \Vxi$. In this way,
\begin{equation}
\label{conC}
\hat{A} = \frac{1}{(2\pi\hbar)^N}\int d\Vxi \;
\tilde{A}(\Vxi)\; \hat{T}_{\Vxi} \
\end{equation}
and the expansion coefficient, a function on $\bf{R}^{2N}$, is the
{\it chord symbol} of the operator $\hat{A}$:
\begin{equation}
\label{covC}
\tilde{A}(\Vxi) = {\rm tr}
\;(\hat{T}_{-\Vxi}\;\hat{A})  .
\end{equation}

The Fourier transform of the translation operators defines the
{\it reflection operators},
\begin{equation}
2^N\;\hat{R}_\x = \frac{1}{(2\pi\hbar)^N}\int d\Vxi \;
\exp{\{\frac{i}{\hbar}(\x\wedge\Vxi)\}} \; \hat{T}_{\Vxi}  ,
\label{refl}
\end{equation}
such that each of these corresponds classically to a reflection of
phase space $\bf{R}^{2N}$ through the point $\x$, that is $\x_0
\mapsto 2\x - \x_0$. The same operator $\hat{A}$ can then be
decomposed into a linear superposition of reflection operators
\begin{equation}
\label{conW} \hat{A} = 2^N\int \frac{d\x}{(2\pi\hbar)^N}\; A(\x)
\; \hat{R}_{\x} ,
\end{equation}
thus defining the {\it centre symbol or Weyl symbol} of operator
$\hat{A}$ \cite{Grossmann},
\begin{equation}
\label{covW}
A(\x) = 2^N{\rm tr}\;(\hat{R}_{\x}\;\hat{A}) .
\end{equation}
It follows that the centre and chord symbols are always related by
Fourier  transform:
\begin{equation}
\tilde{A}(\Vxi) = \frac{1}{(2\pi\hbar)^N}\int d\x \;A(\x)
\exp{\{\frac{i}{\hbar}(\x\wedge \Vxi)\}} \ ,
\label{FWS}
\end{equation}
\begin{equation}
A(\x) = \frac{1}{(2\pi\hbar)^N}\int d\Vxi\; \tilde{A}(\Vxi)
\exp{\{\frac{i}{\hbar}(\Vxi\wedge \x)\}} \ .
\label{FCS}
\end{equation}

In the case of the density operator, $\hat{\rho}$, it is
convenient to normalize its chord symbol, so that we define the
{\it chord function} as
\begin{equation}
\label{chordfunction}
\chi(\Vxi) = \frac{1}{(2\pi\hbar)^N}{\rm tr}
\;(\hat{T}_{-\Vxi}\;\hat{\rho}) =
\frac{\tilde{\rho}(\Vxi)}{(2\pi\hbar)^N} \ ,
\label{trxi}
\end{equation}
whose Fourier transform is the {\it Wigner function},
\begin{equation}
W(\x) = \frac{1}{(2\pi\hbar)^N}\int d\Vxi
\exp{\{\frac{i}{\hbar}(\Vxi\wedge \x)\}}\; \chi(\Vxi),
\label{FTW}
\end{equation}
or alternatively \cite{Royer}
\begin{equation}
W(\x) = \frac{1}{(\pi\hbar)^N}{\rm
tr}\;(\hat{R}_{\x}\;\hat{\rho}).
\label{Wtr}
\end{equation}
The expectation value of any operator $\hat{A}$, defined as
\begin{equation}
\langle \hat{A} \rangle = {\rm tr}\;(\hat{\rho}\;\hat{A}),
\end{equation}
can then be written, according to (\ref{conW}),
\begin{equation}
\langle \hat{A} \rangle = 2^N\int \frac{d\x}{(2\pi\hbar)^N}\;
A(\x) \; {\rm tr}\;(\hat{\rho}\; \hat{R}_{\x}) \ = \int d\x\;
A(\x) \; W(\x) \ ,
\end{equation}
which justifies the Wigner function being dubbed a
``quasi-probability'', even though it can be negative. The
normalization condition reads
\begin{equation}
1 = {\rm tr}\;\hat{\rho} = \int d\x \; W(\x) = (2\pi\hbar)^N
\chi(\vct 0) \ .
\label{normalization}
\end{equation}
The Weyl representation and its Fourier transform have a long
history. References \cite{Grossmann, Royer, Weyl, Wigner, Groen, 
Moyal, Ber77, BalazsJen} develop many of its aspects, with
unavoidable variations in notation and interpretation. Our
presentation is largely based on the review \cite{Report}.

Standard quantum mechanical treatments requires us to choose
between representations based on conjugate variables. This is just
as true for the centre and chord symbols, related by (\ref{FWS})
and (\ref{FCS}), as for the more familiar position and momentum
representations. However, the WKB semiclassical treatment links
the $\x$ variable and the $\Vxi$ variable through the stationary
phase approximation. Indeed, starting from the integral
expressions of (\ref{Wtr}) or (\ref{chordfunction}), this
stationary phase method replaces respectively an integral over
$\x$, or $\Vxi$, by its integrand, evaluated at one or several
points $\x_i$, or $\Vxi_j$. Because of the Fourier relation
(\ref{FTW}) between the pair of representations, each chord $\Vxi$
is then associated with a discrete set of ``centres'' $\x$ - this
denomination will become clear in the following - while each
``centre'' $\x$ specifies a discrete set of chords $\Vxi$.

This correspondence is geometrically clear in the case of a pure
state, $\hat{\rho}_\psi=|\psi\rangle \langle\psi|$, classically
associated with a (quantized) lagrangian submanifold,
$\mathcal{L}_\psi$, in the simple phase space $\x\in\bf{R}^{2N}$,
that is, an $N$-dimensional submanifold $\mathcal{L}_\psi$ with
the property that \begin{equation} \oint_{\gamma} \vct p\cdot \rmd \vct q = 0,
\label{lagr} \end{equation} for any reducible circuit $\gamma$ lying in
$\mathcal{L}_\psi$ (see, e.g. \cite{Weinstein, Arnold, AbMarsden},
for more on symplectic manifolds and their lagrangian
submanifolds). Then, for every point $\x$, one can draw a discrete
set of chords $\Vxi_j$ of the submanifold $\mathcal{L}_\psi$, such
that $\Vxi_j = \x_j^+ - \x_j^-$ and $\x$ is the midpoint of
$[\x_j^-,\x_j^+]$. Reciprocally, every vector $\Vxi$ coincides
with a discrete set of chords for $\mathcal{L}_\psi$, with their midpoints
at $\x_j$. These are the basic elements for the construction of a WKB
semiclassical theory of density operators using this pair of
conjugate representations, as was first noticed by Berry
\cite{Ber77}.

More explicitly, the construction of chords from centres, or vice
versa, is realized as follows: To determine the set of centres
that are conjugate to a given chord, $\Vxi$, for each
$\hat{\rho}$, first translate the whole lagrangian submanifold,
$\mathcal{L}$, by the vector $-\Vxi$, then pick the set
$\{{\x^-}_j\}$ of all points of intersection between $\mathcal{L}$
and the translated submanifold $\mathcal{L}_{-\Vxi}$. The midpoint
of each straight line, between ${\x^+}_j= {\x^-}_j+ \Vxi$ and
${\x^-}_j$, defines $\x_j(\Vxi) = {\x^-}_j + \Vxi /2$, the centre
associated to $\Vxi$ \cite{AlmVal04, emaranha}. To determine set
of chords associated to each centre $\x$, first reflect
$\mathcal{L}$ through $\x$ and pick the set $\{{\x^\pm}_j\}$ of
all points of intersection between $\mathcal{L}$ and the reflected
submanifold $\mathcal{L}_\x$. Then, each reflected pair of
intersections defines a chord associated to $\x$
\cite{AlmHan82,emaranha}, i.e. $\Vxi_j(\x)={\x^+}_j-{\x^-}_j$.

Given $\hat{\rho}$ and the corresponding $\mathcal{L}$, the
simplest semiclassical approximation for the chord function
$\chi(\Vxi)$ relates an amplitude $\alpha_j(\Vxi)$ and a phase
${\sigma}_j(\Vxi)$ to each of the centres $\x_j(\Vxi)$ above, so
that \cite{AlmVal04}
\begin{equation}
\label{semichord}
\chi(\Vxi) = \sum_j \;
\alpha_j(\Vxi)\;\e^{i{\sigma}_j(\Vxi)/\hbar} = \sum_j \;
\chi_j(\Vxi)\ ,
\end{equation}
in such a way that
\begin{equation}
\x_j(\Vxi) =  \J \frac{\partial \sigma_j}{\partial \Vxi} \ .
\label{centrederiv}
\end{equation}
Similarly, the simplest WKB semiclassical approximation for the
Wigner function \cite{Ber77}
\begin{equation}
\label{semiwigner}
W(\x) = \sum_{j}\; a_j(\x) \;\e^{i
S_j(\x)/\hbar} = \sum_j \; W_j(\x) \ ,
\end{equation}
relates an amplitude $a_j(\x)$ and a phase $S_j(\x)$  to each of
the chords $\xi_j(\x)$ above, in such a way that
\begin{equation}
\Vxi_j(\x) =  -\J \frac{\partial S_j}{\partial \x} \ .
\label{chorderiv}
\end{equation}

The phases ${\sigma}_j(\xi)$ (or $S_j(\x)$) are also specified
geometrically by half the action (or symplectic area) of a
circuit taken along the original submanifold $\mathcal{L}$ and
closed along the translated submanifold $\mathcal{L}_{-\Vxi}$ (or
the reflected submanifold $\mathcal{L}_\x$). \footnote{Further
{\it Maslov corrections} \cite{Maslov} should be included in the
phase of the WKB semiclassical Wigner functions \cite{Ber77} and
chord functions \cite{AlmVal04}. These are semiclassically small
and do not alter the geometric relations (\ref{chorderiv}) and
(\ref{centrederiv}).} The fact that the possible chords associated
to a given centre always come in pairs ($\pm \Vxi_j$) guarantees
that the semiclassical Wigner function is real, as it should be.
There is no such restriction for the chord function, unless the
manifold itself has a special symmetry \cite{AlmVal04}.

This simplest semiclassical approximation for the chord and Wigner
functions is valid far from caustics, which arise for arguments of
the chord function whose associated centres coalesce, or for
arguments of the Wigner function whose associated chords coalesce,
respectively. Hence, caustics are related to points of tangency
between $\mathcal{L}$ and $\mathcal{L}_{-\Vxi}$, or between
$\mathcal{L}$ and $\mathcal{L}_\x$, respectively \cite{Ber77,
AlmHan82, AlmVal04, Zambrano}. For the Wigner function, this occurs whenever
$\x$ approaches $\mathcal{L}$, in which case every pair of
associated chords coalesce at a null chord (however, $\mathcal{L}$
is not the only region of centre caustics, generically). The null
chord caustic is more severe for the chord function, because in
this case the entire manifolds $\mathcal{L}$ and $\mathcal{L}_{-\Vxi}$
coincide. Thus, all the points in $\mathcal{L}$ 
are associated centres to the null chord.

The amplitude of each term in the above semiclassical
approximation depends on $N$ variables that are constant along
$\mathcal{L}$. Defining the initial quantum state as an eigenstate
of $N$ commuting quantum operators, the corresponding lagrangian
surface, $\mathcal{L}$ (an $N$-dimensional torus, if it is
compact) will be defined by $N$ action variables
$\mathcal{I}_n(\x)$ in involution,  i.e. all the Poisson brackets
$\{\mathcal{I}_n, \mathcal{I}_{n'}\}=0$. Let us now define the
transported action variables, 
\begin{equation}
\mathcal{I}_n^\pm=\mathcal{I}_n(\x^\pm)=\mathcal{I}_n(\x\pm \Vxi/2),
\label{transaction1} 
\end{equation} 
which may be resolved into either a function of $\x$, for fixed $\Vxi$, 
or vice versa. Then, generally, $\{\mathcal{I}_n^+,
\mathcal{I}_{n'}^-\}\neq 0$ and it is found that the amplitudes
are \begin{equation} a(\x)= |\det \{\mathcal{I}_n^+,
\mathcal{I}_{n'}^-\}|^{-1/2} =\alpha(\Vxi), \label{amplitude} \end{equation}
within an overall normalization constant. This determinant can be
reexpressed in terms of the Jacobian between the centre or chord
variable and the $2N$ variables $(\mathcal{I}_n^+,
\mathcal{I}_{n'}^-)$ \cite{AlmHan82}: \begin{equation} \big|\det
\frac{\partial(\mathcal{I}_n^+,
\mathcal{I}_{n'}^-)}{\partial\x}\big|= |\det\{\mathcal{I}_n^+,
\mathcal{I}_{n'}^-\}|= \big|\det \frac{\partial(\mathcal{I}_n^+,
\mathcal{I}_{n'}^-)}{\partial\vct\xi}\big|. \label{ampl2} \end{equation}
Clearly, the amplitudes, $\alpha_j(\Vxi)$ (or $a_j(\x)$), depend
on the degree of transversality of the intersection between
$\mathcal{L}$ and $\mathcal{L}_{\Vxi}$ (or $\mathcal{L}$ and
$\mathcal{L}_\x$) and so they diverge at caustics \cite{Ber77,
AlmHan82, AlmVal04, Zambrano}.

It should be noted that the equality between the amplitudes in
both representations, equations (\ref{amplitude}) and
(\ref{ampl2}), holds for a specific pair of points $(\x^-,\x^+)$
on the torus and hence for a specific centre-chord pair. In the
centre representation, the Poisson brackets are considered as
functions of $\x$ and we define $\x^\pm(\x)$. For the chord
representation, these same endpoints are a function of $\Vxi$ and
so are the above Poisson brackets. The index, $j$, for the branch
of the chord function (or the Wigner function) has been ommited from
(\ref{amplitude}), because a specific centre-chord pair $(\x,\;
\Vxi)$ will be a particular member of a set $\{(\x,\;
\Vxi_j(\x))\}$ for the Wigner function and, generically, a member
of another set $\{(\x_{j'}(\Vxi),\; \Vxi)\}$ for the chord
function.

\section{Review of the semiclassical limit for unitary evolution}

A theory for the semiclassical limit of unitary evolution,
appropriate to density operators or unitary operators in closed
systems, has been established in both Weyl and chord
representations \cite{RiosOA, OsKon, AlmBro06}. It is worthwhile
to adapt the deduction of phase space propagators in
\cite{AlmBro06} for the needs of the foregoing theory. The
starting point is the product formula for any pair of operators,
$\opB\opA$, in the chord representation:
\begin{equation}
\label{product1} (\opB\opA)(\Vxi) = \frac{1}{(2\pi\hbar)^N}\int
d\xi'\; \tilde{A}(\Vxi')\;\tilde{B}(\Vxi-\Vxi')\;
e^{\frac{i}{2\hbar}(\Vxi\wedge\Vxi')}
\end{equation}
(see e.g. \cite{Report}). Here, when dealing with products of
operators, we abuse the notation and use $(\opB\opA)(\Vxi)$ to
denote the chord symbol $\tilde{C}(\Vxi)$ of the operator
$\widehat C =\opB\opA$ and, similarly, $(\opB\opA)(\x)$ stands for
the Weyl symbol $C(\x)$ 
\footnote{Sometimes, the Weyl symbol of
$\opB\opA$ is denoted by the {\it star product} $B\star A$ , when $B$ is the
Weyl symbol of $\opB$ and $A$ is the Weyl symbol of $\opA$.}.

The problem is that we will work with the chord representation of
$\oprho$, though the Hamiltonian should be specified in the Weyl
representation. This latter is indeed a smooth function, $H(\x)$,
exactly classical, or at least close to it within the order of
$\hbar^2$, whereas its Fourier transform, $\tilde{H}(\Vxi)$, is
highly singular. By defining the translation of an operator as 
\begin{equation}
\hat{A}_{\vct \eta} := \hat{T}_{\vct \eta}\hat{A}\hat{T}_{-\vct
\eta}  , \label{transl} 
\end{equation} 
whose chord representation is given by
\begin{equation}
\label{transchord} \tilde{A}_{\vct \eta}(\Vxi) =
e^{\frac{i}{\hbar}\vct \eta\wedge\Vxi}\; \tilde{A}(\Vxi)  ,
\end{equation}
while its Weyl representation reads
\begin{equation}
\label{transweyl} A_{\vct \eta}(\x) = A(\x + \vct \eta)  ,
\end{equation}
the phase factor in (\ref{product1}) can be incorporated into an
integral involving both representations. Then, using (\ref{refl}),
(\ref{transchord}) and (\ref{transweyl}), we rewrite
(\ref{product1}) as
\begin{equation}
\label{product2} (\opB\opA)(\Vxi) = \frac{1}{(2\pi\hbar)^{2N}}\int
d\Vxi' d\x' \; A(\x'-\Vxi/2) \; \tilde{B}(\Vxi') \;
e^{\frac{i}{\hbar}\x'\wedge(\Vxi-\Vxi') } .
\end{equation}

In this way, we obtain the chord representation of the commutator
between $\hat{H}$ and the evolving density operator,
$\hat{\rho}(t)$, as the {\it mixed} integral,
\begin{equation}
\label{comm} 
\fl (\hat{H}\hat{\rho}-\hat{\rho}\hat{H})(\Vxi)= \int
\frac{d\Vxi' d\x'}{(2\pi\hbar)^{ N}}\;
[H(\x'+\Vxi/2)-H(\x'-\Vxi/2)] \; \chi(\Vxi')\;
e^{\frac{i}{\hbar}\x'\wedge(\Vxi-\Vxi') }.
\end{equation}
Here, we emphasize, $H(\x)$ is the Weyl representation of $\opH$,
which is a smooth function, so that (\ref{comm}) can be integrated
in the stationary phase approximation. In the special case where
$H(\x)$ is a polynomial, we can perform the integrals in
(\ref{comm}) exactly. For a quadratic Hamiltonian we thus
re-derive \cite{BroAlm04} 
\begin{equation} (2\pi\hbar)^{-N}\;
(\hat{H}\hat{\rho}-\hat{\rho}\hat{H})(\Vxi)=
i\hbar\;\Big\{H(\Vxi),\; \chi(\Vxi)\Big\}, 
\end{equation} 
which emulates the
familiar result that the Wigner function evolves classically,
when the Hamiltonian is quadratic \cite{Moyal}.

For general Hamiltonians, we now insert the semiclassical
approximation (\ref{semichord}) for $\chi(\Vxi,t)$ in
(\ref{comm}). Because of the linearity of the evolution equation
for the density operator, it can be decomposed into branches
$\oprho_j(t)$, each evolving separately, as represented by one of
the semiclassical components, $\chi_j(\Vxi, t)$ in
(\ref{semichord}). Then (\ref{comm}) can be integrated by
stationary phase, to yield the lowest order semiclassical
approximation:
\begin{eqnarray}
\nonumber
(2\pi\hbar)^{-N}\;(\hat{H}\hat{\rho}-\hat{\rho}\hat{H})_{SC}(\Vxi)\\
=\sum_j\alpha_{j}(\Vxi)
\Big(H\big(\J\frac{\partial\sigma_{j}}{\partial\Vxi} +
\frac{\Vxi}{2}\big) -H\big(\J\frac{\partial\sigma_{j}}{\partial
\Vxi} -
\frac{\Vxi}{2}\big)\Big)\;e^{i\sigma_{j}(\Vxi)/\hbar} \\
= \sum_j \Big(H(\x_{j}(\Vxi) +\Vxi/2) - H(\x_{j}(\Vxi) -
\Vxi/2)\Big)\;\chi_{j}(\Vxi) \ . \label{SCcomm}
\end{eqnarray}
Thus, by comparing with the unitary part of the master equation
(\ref{Lindblad}), we find that the classical chord action
$\sigma_{j}(\Vxi,t)$ evolves according to the Hamilton-Jacobi
equation \cite{AlmBro06}:
\begin{equation}
-\frac{\partial \sigma_{j}}{\partial t}(\Vxi,t) =
H\big(\J\frac{\partial \sigma_{j}}{\partial \Vxi} +
\frac{\Vxi}{2}\big) - H\big(\J\frac{\partial
\sigma_{j}}{\partial\Vxi} - \frac{\Vxi}{2}\big), \label{HJ_chord}
\end{equation}
similarly to the evolution for the centre action \cite{OsKon}.

It must be remembered that in (\ref{HJ_chord}), as well as in
(\ref{comm}) through (\ref{SCcomm}), the function $H(\x)$ is the
Weyl representation of the quantum hamiltonian operator, which
will be either identical, or semiclassically close to the
classical hamiltonian function. In the general case where this
hamiltonian function is nonlinear, the resulting evolution of the
chord action is a consequence of the classical motion
$\x_j^{\pm}(\Vxi,t)$  of both the initial chord tips, 
$\x_j^{\pm}(\Vxi,0)= \x_j(\Vxi) \pm \Vxi/2$, whereas neither the
chord, $\Vxi$ itself, nor the corresponding centres, $\x_j(\Vxi)$,
will generally follow their respective hamiltonian phase space
trajectories \cite{RiosOA}. By working directly on the double
phase space, as discussed in section 6, a new Hamiltonian function
can be defined on this doubled space to take account of the motion
of both chord tips in a single trajectory \cite{OsKon}.

The above approximation for the unitary evolution of the chord
function does not include the evolution of the amplitudes,
$\alpha_j(\Vxi)$ in (\ref{semichord}), which can be obtained by
including the next order in $\hbar$ in the theory. Alternatively,
we note that, up to the leading order, the evolution can be
portrayed as resulting from the full classical motion, i.e. all
the trajectories generated by the hamiltonian, $H(\x)$, which
transports the entire lagrangian submanifold, $\mathcal{L}(t)$,
and its neighborhood. Thus, each pair of points, $\x_j^{\pm}(\Vxi,
t)$ on $\mathcal{L}(t)$ defines an evolving chord, 
\begin{equation}
\overline{\Vxi}_j(t)=\x_j^{+}(\Vxi, t)-\x_j^{-}(\Vxi,t)
\label{chorddyn} 
\end{equation} 
and an evolving centre, 
\begin{equation}
\overline{\overline{\x}}_j(\Vxi , t)=\Big(\x_j^{+}(\Vxi,
t)+\x_j^{-}(\Vxi, t)\Big)/2 \ . 
\label{centerdyn} 
\end{equation} 
One should note that, here, 
$\x_j^{\pm}(\Vxi, t)$ denotes the hamiltonian trajectories of
$\x_j^{\pm}(\Vxi, 0)= \x_j^{\pm}(\Vxi) = \x_j(\Vxi) \pm \Vxi/2$.
Therefore, $\overline{\Vxi}_j(t)$ is generally different from the
hamiltonian trajectory $\Vxi_j(t)$ of the initial chord,
$\Vxi_j(0)=\overline{\Vxi}_j(0)=\Vxi$, 
unless the hamiltonian is quadratic. Similarly,
$\overline{\overline{\x}}_j(\Vxi,t)$ is generally different from
the hamiltonian trajectory $\x_j(\Vxi,t)$ of
$\x_j(\Vxi,0)=\overline{\overline{\x}}_j(\Vxi,0)=\x_j(\Vxi)$
\cite{RiosOA}.

By reconstructing the chord function according to the
semiclassical prescription (\ref{semichord}) at each instant, the
same phase evolution is obtained as from the Hamilton-Jacobi
equation (\ref{HJ_chord}), but now the evolution of the amplitudes
will also be included, as long as we also allow the action
variables $\mathcal{I}_n(\x^{\pm})$ in (\ref{amplitude}) to evolve
according to $\mathcal{I}_n(\x^{\pm},
t)=\mathcal{I}_n(\x^{\pm}(t))$, where $\x^{\pm}(t)$ is the
hamiltonian trajectory of $\x^{\pm}(0)$. Again, it must be
stressed that this semiclassical evolution of the chord function
(or the Wigner function), resulting from global classical motion
together with the geometric reconstruction of the representation
at each instant, can only be identified with Liouville evolution
(i.e. the evolution obtained from the hamiltonian trajectory of the
argument of the the chord function, or the Wigner function) if
the Hamiltonian is quadratic \cite{RiosOA}.

\section{Chord representation of the open interaction term}

We now address various integral representations of the chord
symbol for the open interaction term. The starting point is the
product rule in the chord representation \cite{Report}, 
\begin{equation} 
\fl (\opA \opB \widehat C)(\Vxi) =\!\!\int
\frac{d\Vxi'd\Vxi''d\Vxi'''}{(2\pi\hbar)^{2N}}
\tilde{A}(\Vxi')\tilde{B}(\Vxi'')\tilde{C}(\Vxi''')\;
\delta(\Vxi-\Vxi'-\Vxi''-\Vxi''') \; \exp\left[\frac{i}{2\hbar}
(\Vxi\wedge\Vxi'-\Vxi''\wedge\Vxi''')\right], 
\label{product3a}
\end{equation} 
where, again, we abuse the notation and write $(\opA \opB
\widehat C) (\Vxi)$ for the chord symbol of $\opA \opB \widehat
C$. The exponent in the integrand is here one of the many
different expressions for the symplectic area of the quadrilateral
with sides: $\Vxi', \Vxi'', \Vxi''', -\Vxi$. Incorporating the
phase factor for translation into the chord representation, as in
(\ref{transchord}), leads to the compact expression, 
\begin{equation} 
(\opA\opB \widehat C)(\Vxi)=\int
\frac{d\Vxi'd\Vxi''}{(2\pi\hbar)^{2N}}\;
\tilde{A}_{\Vxi/2}(\Vxi')\;\tilde{B}(\Vxi'')\;
\tilde{C}_{-\Vxi''/2}(\Vxi-\Vxi'-\Vxi''). 
\label{product3} 
\end{equation} 
Even if the Lindblad operators $\hat{L}$ are not observables, as in the
optical example (2), their Weyl representation are smooth
functions on phase space, $L(\x)$, whereas their chord
representation, $\tilde{L}(\Vxi)$, are quite singular. Therefore,
we again need a mixed product rule, where a pair of operators,
$\opA$ and $\widehat C$ are expressed in the Weyl representation:
\begin{equation} 
\fl (\opA \opB \widehat C)(\Vxi)=\int
\frac{d\x'd\Vxi''}{(2\pi\hbar)^{2N}}\;
A_{\Vxi/2}(\x')\;\tilde{B}(\Vxi'')\;C_{-\Vxi''/2}(\x')\;
\exp\left[\frac{i}{\hbar}\x'\wedge(\Vxi-\Vxi'')\right] .
\label{product4} 
\end{equation}

To obtain the desired expression for the nonunitary term of the
Lindblad equation (\ref{Lindblad}), the order of the operators is
permuted, which leads to sign changes for translated operators
(\ref{transl}), given by $A_{\vct \eta}(\x) = A(\x + \vct \eta)$
in the Weyl representation, so that
\begin{eqnarray}
\fl (\hat{L}\hat{\rho}\hat{L}^{\dag} -
\frac{1}{2}\hat{L}^{\dag}\hat{L}\hat{\rho} -
\frac{1}{2}\hat{\rho}\hat{L}^{\dag}\hat{L})(\Vxi) = \int
\frac{d\Vxi'd\x'}{(2\pi\hbar)^{N}}  \nonumber \ \ \chi(\Vxi')\;
\exp{\Big[\frac{i}{\hbar}\x'\wedge(\Vxi-\Vxi')\Big]} \nonumber\\
\fl \Big\{\! L(\x'\! + {\Vxi\over 2})L(\x'\! -{\Vxi'\over 2})^{*}
\!-\!\frac{1}{2}\big[L(\x'\! + {\Vxi'\over 2})L(\x'\! +
{\Vxi\over 2})^{*} + L(\x'\! -{\Vxi\over 2})L(\x'\! - {\Vxi'\over
2})^{*}\big]\Big\}. \label{L7}
\end{eqnarray}
Note that $L(\x'\! + {\Vxi/2})^{*}$ is the Weyl symbol of the
operator $\hat{L}^{\dag}$ translated by ${\Vxi/2}$, which is
not equal to the adjoint of $\hat{L}_{\Vxi/2}$.

The exact formula (\ref{L7}) is at a par with the representation
of the commutator (\ref{comm}). It is interesting that, although
(\ref{L7}) represents products of three operators, the dimension
of the integral is the same as in (\ref{comm}). Thus, including
the presence of an internal Hamiltonian and a single Lindblad
operator, the exact equation of motion for the chord function is
given by
\begin{eqnarray}
 \fl \hbar \frac{\partial \chi}{\partial t} (\Vxi,t) =
\int \frac{d\Vxi'd\x'}{(2\pi\hbar)^{2N}} \;\; \chi(\Vxi',t) \;
\exp{\Big[\frac{i}{\hbar}[\x'\wedge(\Vxi-\Vxi')]\Big]}\;\;
\Big\{-{\rm i}[H(\x' - {\Vxi\over 2}) - H(\x' + {\Vxi\over 2})]  \nonumber \\
\fl \;\;+\;  
\Big[  L(\x' + {\Vxi\over 2})L(\x'-{\Vxi'\over 2})^{*}
-\frac{1}{2} \big[L(\x' +{\Vxi'\over 2})L(\x' + {\Vxi\over
2})^{*} +L(\x' - {\Vxi\over 2}) L(\x' -{\Vxi'\over
2})^{*}\big]\Big]\Big\}. \label{chordyneq}
\end{eqnarray}
If there are more Lindblad operators in the master equation
(\ref{Lindblad}), then one must sum over these in the integrand on
the right hand side of (\ref{chordyneq}). We have not included
this obvious extension, so as not to confuse this sum with the
further sum over semiclassical branches in the following formulae.

Far from caustics, one can evaluate (\ref{L7}) approximately, 
by stationary phase, if $L(\x)$ is assumed to be a smooth function, 
by inserting the semiclassical approximation 
for each separate branch of the chord function
(\ref{semichord}) as in the previous section. The stationary phase
condition singles out $\Vxi'=\Vxi$ and $\x'=\x_j(\Vxi)$, one of
the centres associated to a geometrical chord $\Vxi$ of the
classical submanifold $\mathcal{L}$. The full semiclassical
approximation is simply
\begin{eqnarray}
\fl (2\pi\hbar)^{-N}(\hat{L}\hat{\rho}\hat{L}^{\dag} -
\frac{1}{2}\hat{L}^{\dag}\hat{L}\hat{\rho} -
\frac{1}{2}\hat{\rho}\hat{L}^{\dag}\hat{L})_{SC}(\Vxi) = \sum_j
\Big\{L(\x_j(\Vxi) + \Vxi/2)L(\x_j(\Vxi)-\Vxi/2)^{*}
\nonumber\\
-\frac{1}{2}\big\{|L(\x_j(\Vxi) + \Vxi/2)|^2 + |L(\x_j(\Vxi)
-\Vxi/2)|^2\big\}\Big\}\;\alpha_j(\Vxi)\;\;
e^{i\sigma_j(\Vxi)/\hbar}.
\end{eqnarray}
In terms of the chord tips, $\x_j^{\pm}(\Vxi) = \x_j(\Vxi) \pm
\Vxi/2$, the semiclassical approximation to the chord
representation of the open interaction term can be rewritten as
\begin{eqnarray}
(2\pi\hbar)^{-N}(\hat{L}\hat{\rho}\hat{L}^{\dag} -
\frac{1}{2}\hat{L}^{\dag}\hat{L}\hat{\rho} -
\frac{1}{2}\hat{\rho}\hat{L}^{\dag}\hat{L})_{SC}(\Vxi) = \nonumber \\
\fl\sum_j \Big\{ -\frac{1}{2} |L(\x_j^+(\Vxi)) -
L(\x_j^-(\Vxi))|^2 + {\rm
i}\;\mathcal{I}m\{L(\x_j^+(\Vxi))L(\x_j^-(\Vxi))^{*}\}\Big\}\;\chi_j(\Vxi),
\label{P9}
\end{eqnarray}
where $\mathcal{I}m$ denotes the imaginary part and $\chi_j(\Vxi)$
is a branch of the semiclassical chord function given by
(\ref{semichord}). In the case of a linear function,
\begin{equation}
 L(\x) = \Vl\cdot \x = \Vl'\cdot \x + i \;\Vl''\cdot \x \ ,
\label{lin}
\end{equation}
as in the optical example (\ref{opt}), the semiclassical
approximation for the open interaction term simplifies to
\begin{eqnarray}
 (2\pi\hbar)^{-N}(\hat{L}\hat{\rho}\hat{L}^{\dag} -
\frac{1}{2}\hat{L}^{\dag}\hat{L}\hat{\rho} -
\frac{1}{2}\hat{\rho}\hat{L}^{\dag}\hat{L})_{SC}(\Vxi) = \nonumber \\
\sum_j \left(i(\Vl'\wedge \Vl'') \x_j(\Vxi)\wedge \Vxi-{1\over
2}[(\Vl'\cdot\Vxi)^2+(\Vl''\cdot\Vxi)^2]\right) \chi_j(\Vxi).
\label{semilin}
\end{eqnarray}

On the other hand, (\ref{L7}) can be integrated exactly, for a
linear Lindblad operator (or even if it is a polynomial). Then
(\ref{L7}) becomes
\begin{equation}
\fl (2\pi\hbar)^{-N}(\hat{L}\hat{\rho}\hat{L}^{\dag} -
\frac{1}{2}\hat{L}^{\dag}\hat{L}\hat{\rho} -
\frac{1}{2}\hat{\rho}\hat{L}^{\dag}\hat{L})(\Vxi) = \hbar
(\Vl'\wedge\Vl'') \; \Vxi\cdot\frac{\partial \chi}{\partial \Vxi}
- \frac{1}{2}\;\Big[(\Vl'\cdot\Vxi)^2 + (\Vl''\cdot\Vxi)^2\Big] \;
\chi(\Vxi) \ , \label{exactdiss}
\end{equation}
in agreement with \cite{BroAlm04}. Compared with (\ref{semilin}),
we find the same second term on the right hand side. If the
Lindblad operator is self-adjoint, i.e. $\Vl''=\vct 0$, this will
be the only term. In this case, it is easier to develop a
semiclassical theory for evolution of the density operator, which
becomes exact in the case that the Hamiltonian is quadratic. This
will be pursued in the following section. The first term was shown
to describe dissipation in the exact quadratic theory
\cite{BroAlm04}. Though dissipation cannot be included in a standard
semiclassical theory, we will show that it is naturally
accommodated within the double phase space formalism that is
developed in later sections.

\section{Decoherence without dissipation}

In this section, all Lindblad operators $\hat{L}_k$ are restricted
to be self-adjoint, so that
$\mathcal{I}m\{L(\x_j^+(\Vxi))L^*(\x_j^-(\Vxi))\} \equiv 0$,
simplifying equation (\ref{P9}). As pointed out in the
introduction, this means that the system may be considered to be
conservative, albeit open to a random environment.

If we further ignore the internal hamiltonian motion, or, more
reasonably, restrict analysis of the decoherence process to its
first stages, then we can consider the action $\sigma_j(\Vxi)$ to
be constant in time, while the semiclassical amplitude evolves as
\begin{equation}
\label{amplichordt} \alpha_j(\Vxi,t) =
\alpha_j(\Vxi,0)\exp{\Big\{-\frac{t}{2\hbar}\sum_k\big|L_k(\x_j^+(\Vxi))
- L_k(\x_j^-(\Vxi))\big|^2\Big\}} \ .
\end{equation}
Generally, the above equation implies a fast shrinking of the
chord function to a progressively narrower neighborhood of the
origin. According to the discussion in \cite{BroAlm04}, this
accounts for a fast loss of quantum correlations. However, for
those chord tips, $\x_j^{\pm}$, that lie on a level curve (or
level surface) of one of the real functions, $L_k(\x)$, this term will
not contribute to the loss of amplitude. The condition for a
chord not to decay at all is that its tips should lie on the
intersection of level surfaces for all the functions, $L_k(\x)$.

The effect of the internal Hamiltonian, $\hat{H}$, can be included
by considering the limit in a process where we switch it on and
off, while alternatively connecting and disconnecting the Lindblad
interaction (opening and closing the system)  \cite{OA03}. 
This defines a periodic markovian system in the limit of small
periods, as in the periodization of hamiltonian systems in
\cite{Berryetal}. In the limit of short periods, there results a generalized 
{\it Trotter ansatz} \cite{Schulman}.
\footnote{One should note that the proof of the Trotter theorem
does not require full groups, appropriate for unitary evolution,
but also encompasses {\it semigroups}, as in the present case.}
Both the tips of the chord, $\x_j^{\pm}(\Vxi)$,
will evolve classically as $\x_j^{\pm}(\Vxi,t)$ according to the
Hamilton-Jacobi equation (\ref{HJ_chord}) for a time $\tau /2$,
implying in the temporal evolution of a given chord, $\Vxi$, as
$\overline{\Vxi}_j(\Vxi,t)$ and for the centre, $\x_j(\Vxi)$, the
motion $\overline{\overline{\x_j}}(\Vxi,t)$, according to
equations (\ref{chorddyn}) and (\ref{centerdyn}). Then, at each
opening of the system for a further period of $\tau /2$, the
amplitude evolves according to (\ref{amplichordt}). 
Naturally, one must multiply both the open and the closed
terms of the Lindblad equation by a factor of two, to make up
for the reduced time in which either of them acts.
In the limit as $\tau \rightarrow 0$ of an infinite number of {\it
closing} and {\it opening} operations, we obtain the full
semiclassical evolution of the chord function in a region free of
caustics, as $\chi(\Vxi,t)=\sum_j\chi_j(\Vxi,t)$, with
\begin{equation}
\label{chit} \chi_j(\Vxi,t)= {\chi_j}^0(\Vxi,t)\;\;
\exp\Big[\frac{-1}{2\hbar} D\{\x_j^+(-t),\x_j^-(-t)\}\Big],
\end{equation}
where $\x_j^{\pm}(-t)$ is short for $\x_j^{\pm}(\Vxi,-t)$ and
${\chi^0}_j(\Vxi, t)$ denotes the semiclassical propagation for a
time $t$ of the $j$-branch of the chord function for the
corresponding closed system (with all $\widehat L_k =0$). The
decay in amplitude for each branch of the chord function is
determined by the {\it decoherence functional} over trajectory
pairs,
\begin{equation}
D\{\x^+(t),\x^-(t)\} := \sum_k \int_0^t dt'
|L_k(\x^+(t'))-L_k(\x^-(t'))|^2  \ , \label{decfl}
\end{equation}
where $\x^{\pm}(0) = \x(\Vxi)\pm \Vxi/2$. Hence, it is the pair of
backward trajectories ending at a given pair of chord tips on
$\mathcal L(t)$  that determine the decrease in amplitude. The
square root of the decoherence functional is a kind of time
dependent measure of {\it distance} between any pair of points
$(\x^+,\x^-)$, as pointed out by Strunz \cite{Strunz}.

Concerning the derivation of the above semiclassical expression
(\ref{chit}), note that the chord function entering into the
master equation is here the semiclassical chord function of a pure
state (\ref{semichord}), which is valid away from chord caustics.
But far from the origin (a chord caustic), the damping factor in
(\ref{chit}) is a non-oscillatory function. Hence, in a first
approximation, it may be considered as a new factor of the
semiclassical amplitude $\alpha_j(\Vxi)$ in (\ref{chit}), even
though the exponent is divided by $\hbar$. Thus, when this
modified expression for the chord function is inserted into the
master equation, it is still the chord action function from
(\ref{semichord}) that defines the semiclassical evolution of the
decaying chord function (\ref{chit}), as long as all pertinent
integrals are computed via the {\it real} stationary phase
method. Accordingly, an improvement to (\ref{chit}) could in
principle be obtained by computing all pertinent integrals via the
complex steepest descent method. This improvement is at present
being investigated.

The simplest case is where the Lindblad operators are all linear
functions of position and momenta, $L_k(\x)= \Vl_k\cdot \x$. Then
expression (\ref{chit}) simplifies, because
\begin{equation}
\label{linear}
\Big|L_k\big(\x_j^+(\Vxi,-t')\big)-L_k\big(\x_j^-(\Vxi,-t')\big)\Big|^2
= \big|\Vl_k\cdot\overline{\Vxi}_j(\Vxi,-t')\big|^2 \ ,
\end{equation}
where the explicit dependence on $\Vxi$ is emphasized in the
r.h.s. Generally the evolution of each $\Vxi_j$ results from the
hamiltonian flow of the tips $\x_j^{\pm}(\Vxi)$, so that the
evolution is j-dependent. However, if the internal Hamiltonian
is a homogeneous quadratic, then the evolution of the
chord is just given by \cite{BroAlm04} 
\begin{equation}
\dot{\Vxi}=\J\;\frac{\partial H}{\partial \Vxi} 
\end{equation} 
and is therefore $j$-independent. Furthermore, the internal dynamics of
the chord function is then Liouvillian: ${\chi^0}_j(\Vxi, t) =
\chi_j(\Vxi(-t), 0)$. In this way, all $j$-branches of the
semiclassical chord function can be combined into a single
evolution, so that
\begin{equation}
\label{linn} \chi(\Vxi,t)={\chi^0}(\Vxi(-t),0)
\exp{\Big\{-\frac{1}{2\hbar}
\sum_k\int_0^tdt'\big|\Vl_k\cdot\Vxi(-t')\big|^2\Big\}}.
\end{equation}

It is remarkable that this simple expression for the semiclassical
evolution of an open system is actually exact and valid for any
initial chord function (pure or mixed), under the above hypothesis
for $\hat{L}$ and $\hat{H}$ \cite{BroAlm04}. Thus, no matter how
full of quantum correlations the initial state might be, the
infinite product of gaussian exponentials in (\ref{linn}), or the
more general exponential of the decoherence functional in
(\ref{chit}) progressively squeezes them out. This process, 
by which the large chords are quenched, proceeds irreversibly, since
$D\{\x_j^+(-t),\x_j^-(-t)\}$ is a nondecreasing function of time.
The semiclassical expression (\ref{chit}) generalizes the simple
exact solution (\ref{linn}), when the Hamiltonian is not
quadratic, for chords that never lie close to caustics throughout
the evolution. We retain the qualitative
picture in which the evolving chord function is squeezed onto the
origin by the decoherence functional, although
$D\{\x_j^+(t),\x_j^-(t)\}$ is no longer a quadratic function of
$\Vxi$.

The only possibility for the decoherence functional not to increase 
arises for pairs of
classical trajectories generated by $H(\x)$, lying along a level
submanifold of the linear Lindblad-Weyl function, $L(\x)$, 
i.e., the condition is that the Poisson bracket $\{{L},{H}\}=0$, 
which holds when the operators $\widehat L$ and $\widehat H$ commute. 
For more than one Lindblad operator, there is no dampening for those classical
trajectories lying on the intersection of all $L_k$ level
submanifolds, that is, when $\{{L}_k,{H}\}=0$, for all $k$. 
In the quadratic case, the specific evolution for each kind of classical dynamics 
(elliptic, parabolic or hyperbolic) is studied in \cite{BroAlm04}.

The form of the evolution (\ref{chit}) goes some way towards
justifying the rough qualitative description of decoherence (\ref{amplichordt}), 
that neglects the internal dynamics for very short times, since the chord
function is seen to decay exponentially fast in the domain where
(\ref{chit}) is valid (which excludes a neighborhood of the origin).
Generically, $\{{L}_k,{H}\}\neq0$, so that one needs (\ref{chit})
to depict this quenching of the long chords in a fully quantitative manner. 
The further evolution of the state, for longer times,
depends entirely on the remaining small chords, 
so that even our fuller semiclassical description is inappropriate. 
A generalization of the quadratic case (\ref{linn})
into the region of small chords is achieved indirectly in
section 8.

The semiclassical evolution of the Wigner function for open
conservative markovian systems is obtained by the Fourier
transform of the
semiclassical evolution of the chord function. Each term of the
sum, $W(\x,t)=\sum_j W_j(\x,t)$, is given by a convolution
integral of the unitarily evolving branch of the semiclassical
Wigner function (\ref{semiwigner}) with the Fourier transform of
the decaying amplitude term in (\ref{chit}). This diffusive window, which
coarse-grains the Wigner function, will broaden with time, as its
inverse Fourier transform narrows down the range of the chord
function. In the case of a quadratic Hamiltonian, 
the window will be gaussian and this description of the
evolution of the Wigner function becomes exact \cite{BroAlm04}.

The the evolving chord function (\ref{chit}) can only be inserted
into the Fourier transform (\ref{FTW}) for chords that are far
from caustics, which precludes small chords. For large chords, the
convolution integral can be evaluated by (real) stationary phase,
because the decaying amplitude term is a smooth function of $\Vxi$
far from the origin. We then obtain a superposition of terms of
the same form as (\ref{semiwigner}), each of them corresponding to
a different branch of the centre action function $S_j(\x,t)$.
However, as with the chord function, the amplitude $a_j(\x)$ now
acquires a new time-dependent factor, so that we have the complete
analogue of equation (\ref{chit}) for the semiclassical evolution
of the Wigner function as
\begin{equation}
\label{wigt} W_j(\x,t) = {W^0}_{j}(\x,t) \;\; \exp\Big[
\frac{-1}{2\hbar} D\{\x_j^+(-t),\x_j^-(-t)\}\Big],
\end{equation}
where ${W^0}_{j}(\x,t)$ denotes semiclassical propagation for a
time $t$ of the $j$-branch of the Wigner function as a closed
system. The amplitude also decays according to  the decoherence
functional (\ref{decfl}) that quenches the contribution of long
chords. However, this is now determined by the choice of centre,
$\x$, rather than the chord, $\Vxi$, i.e. here the chord tips are
$\x^{\pm}(\x,t)$. One should note that the inclusion of the new
decoherence damping factors in the semiclassical amplitudes 
of the chord function (\ref{chit}) and the Wigner function (\ref{wigt})
does not destroy their equality (\ref{amplitude}).

The semiclassical expression (\ref{wigt}) was derived earlier in
\cite{OA03}, solely within the Weyl representation. However, it
does not lead to the exact evolution of the Wigner function, in the
simple case of linear Lindblad and quadratic Hamiltonian operators,
as does its chord similar (\ref{chit}). It follows that the theory
for markovian open systems does not generalize the customary
semiclassical covariance, among various representations of a given unitary
evolution, with respect to Fourier transforms performed by
stationary phase. Here, the chord representation has a decided
advantage, because an essential qualitative feature is destroyed
if the Fourier transform, leading to the Wigner function, is approximated by
stationary phase.

Thus, although (\ref{wigt}) well describes the initial stages of
decoherence, it fails to address the diffusive process that sets in
at longer time scales. In fact, there is no suggestion of the
{\it decoherence time} threshold at which the initial
pure-state Wigner function becomes positive definite, as always
happens in the case of linear Lindblad and quadratic Hamiltonian
operators \cite{DioKief, BroAlm04}. The specific case of the damped
harmonic oscillator is further analysed in \cite{IsarScheid07}. This
is the time which takes 
an initial state, represented by a Dirac-delta function in phase
space, to evolve into a Gaussian with the width of a pure coherent
state. At this time, any initial pure Wigner function evolves into
a positive-definite phase space distribution, which is
indistinguishable from a Husimi function \cite{Husimi, Takashi}. 
In the general case of nonquadratic Hamiltonians, 
we can still define a local decoherence
time as that which it takes the quenching factor in (\ref{chit})
to shrink to the extent that it has the same area (or volume) as a
coherent state. Beyond this time, its Fourier transform will
coarse-grain away the fine oscillations of the Wigner function.

\section{Semiclassical evolution in double phase space}

WKB theory and its generalization to higher
dimensions \cite{Maslov, Gutzwiller} relates the solution,
$\langle q|\psi(t)\rangle$, of a Schr\"odinger equation to the
corresponding classical Hamiltonian trajectories in the phase
space $\x=(p,q)\in \bf{R}^{2N}$. This Schr\"odinger solution is
the $|q \rangle$ representation of the unitarily evolving state,
associated to a lagrangian submanifold $\mathcal{L}_\psi$, which
is described within the Weyl formalism in section 3. This
submanifold is more commonly described (locally) as a graph of the
classical function, $p(q)=\frac{\der S}{\der q}(q)$, which maps
$q\in\bf{R}^{N}$ onto $p\in\bf{R}^{N}$ (which are also lagrangian
coordinate subspaces, satisfying (\ref{lagr})). The classical
action $S(q)$ is both the generating function for
$\mathcal{L}_\psi$ and the oscillating phase of the quantum wave
$\langle q|\psi\rangle$.

Analogously, the linear operators, $\widehat A$, that act on the
quantum Hilbert space, form a vector space $|A\rangle\!\rangle$,
for which the dyadic operators $|\vct
Q\rangle\!\rangle=|q^-\rangle \langle q^+|$ constitute a complete
basis. Thus, defining the {\it Hilbert-Schmidt product}: 
\begin{equation}
{\rm tr}\>\opA^{\dagger}\widehat B = \langle\!\langle A|B\rangle\!\rangle, 
\label{HJprod} 
\end{equation} 
we can interpret the ordinary position representation 
of the operator $\widehat A$ as
\begin{equation}
\langle q^+|\opA|q^-\rangle= {\rm tr}\>|q^-\rangle\langle q^+|\opA
= \langle\!\langle \vct Q|A\rangle\!\rangle, 
\end{equation}
in close analogy to a wave function.
\footnote{It should be noted that we will not relie on the Hilbert-Schmidt norm
and its evolution in the following discussion.}
It is then natural to relate a {\it double Hilbert space} of
$|ket\rangle\langle bra|$ states to a double phase space:
$\{\X\}=\{\x^-\} \times \{\x^+\}$, where $\x^{\pm} =
(p^{\pm},q^{\pm})$ (see e.g.\cite{Littlejohn95, emaranha}, or,
for non-vectorial cases \cite{RiosOA1}). The operator $|\vct
Q\rangle\!\rangle$ should then correspond to the lagrangian
subspace, $\vct Q=constant$, in the double phase space. This does
hold, within a minor adaptation, due to the presence of the
adjoint operator in the definition of the Hilbert-Schmidt product,
or, more directly, the fact that, in ordinary Hilbert space, 
$\langle bras|$ are adjoint to $|kets\rangle$. 
Accordingly, if we define $\vct Q=(q^-, q^+)$, we
should define $\vct P=(-p^-, p^+)$ as conjugate coordinates on the
double phase space $\{\X=(\vct P,\vct Q)\}$. This is equivalent to
changing the sign of the symplectic structure on ${\bf{R}^{2N}} =
\{\x^-\}$.

In this way, we include, within the set of lagrangian submanifolds
in double phase space, all the graphs of canonical transformations
on single phase space, $\x^- \mapsto \x^+=\C(\x^-)$. That is, we
may rewrite the definition of a canonical transformation as 
\begin{equation}
\oint_{\Gamma} \vct P\cdot \rmd \vct Q = 0, 
\end{equation} 
where $\Gamma$ is
any curve defined on the $(2N)$-dimensional submanifold,
$\Lambda_\C$, which is the graph of the canonical transformation
$\C$ on the $(2N)$-dimensional space $\{\x^- = (q^-,p^-)\}$,
within the $(4L)$-dimensional double phase space, ${\bf{R}^{4N}} =
\{\X=(\vct P,\vct Q)\}$. If $\theta$ is a parameter along $\Gamma$,
then $\Gamma(\theta)=(\gamma^-(\theta),\gamma^+(\theta))$, where $\gamma^-(\theta)
\mapsto \gamma^+(\theta) = \C(\gamma^-(\theta))$, and we may consider the
curves $\gamma^\pm$ as projections of the curve $\Gamma$. Going
back to the operational meaning of this construction, if
$\mathcal{L}^-$ is the lagrangian manifold corresponding to a
quantum state $|\psi^-\rangle$, and $\C$ a canonical
transformation, then $\mathcal{L}^+ = \C(\mathcal{L}^-)$ can be
interpreted as the lagrangian manifold of some $|\psi^+\rangle$
state, and the whole operation corresponds to a unitary quantum
operator, $\opU_\C:|\psi^-\rangle \mapsto |\psi^+\rangle$.

Besides portraying the graph of a canonical transformation as a
Lagrangian submanifold, the product of a Lagrangian submanifold,
$\mathcal{L}^-$ in $\{\x^-\}$ with any another submanifold
$\mathcal{L}^+$ in $\{\x^+\}$, $\Lambda=\mathcal{L}^- \times
\mathcal{L}^+$, is also Lagrangian in double phase space, but
projects singularly onto either of the factor spaces
$\{\x^{\pm}\}$. In the case that both submanifolds are tori, we
obtain a double phase space torus, as if we had doubled the number
of degrees of freedom. If $N=1$, it will be a $2$-dimensional
product torus \cite{emaranha} (taking care with the sign of $p^-$,
in the present construction).

If both Lagrangian submanifolds in single phase space correspond
to the same state, i.e. $|\psi^-\rangle=|\psi^+\rangle$, then we
represent the corresponding pure state density operator,
$\oprho_\psi=|\psi\rangle\langle\psi|=|\Psi\rangle\!\rangle$, in
the $|\vct Q\rangle\!\rangle$ representation as 
\begin{equation}
\langle\!\langle \vct Q|\Psi\rangle\!\rangle=\langle
q^+|\psi\rangle\langle\psi|q^-\rangle. 
\end{equation} 
Therefore, its simplest semiclassical approximation 
can be expressed as a superposition of terms of the form 
\begin{equation} 
\langle\!\langle \vct
Q|\Psi\rangle\!\rangle=A_j(\vct Q) \exp [i \mathcal{S}_j(\vct
Q)/\hbar], 
\end{equation} 
with 
\begin{equation} 
\mathcal{S}_j(\vct Q)=\int_{\vct 0}^{\vct
Q} \vct P_j(\vct Q')\cdot \rmd \vct Q' =\int_0^{q_+} p_j^+\cdot
\rmd q^+ - \int_0^{q_-} p_j^-\cdot \rmd q^-. \label{daction1} 
\end{equation}
Again, this is in strict analogy with the construction of
semiclassical product states of higher degrees of freedom
\cite{emaranha}.

The next step is a change of lagrangian coordinates in double
phase space: \begin{equation} (\vct P,\vct Q) \mapsto (\x,\y) \ , \
\x=\frac{\x^+ + \x^-}{2} \ , \ \y= \J(\x^+ -\x^-)=\J\Vxi.
\label{centrechord} \end{equation} Here, $\J$ is the constant symplectic
matrix in single phase space and is used to {\it canonize} the
initial $\pi/4$ rotation on $(\x^-,\x^+)$ that introduces the pair
of lagrangian coordinates $(\x,\Vxi)$ on double phase space. Thus,
the pair of conjugate variables $(\x,\y)$ also accounts for the
sign change in the $p^-$ coordinate. We should bear the discomfort
that the canonical coordinate in double phase space is $\y$, while
the geometrically meaningful variable in single phase space is
$\Vxi$, the trajectory {\it chord}, which has $\x$ as its {\it
centre}. It would also be possible to choose the variable, $\Vxi$, as
the conjugate to $\x$, instead of $\y$, but at the cost of writing
the symplectic form on double phase space in a noncanonical way,
leading to less familiar expressions for Hamilton's equations and
some other elements of the semiclassical theory (see
\cite{RiosOA1} for some of these expressions).

If we consider the {\it horizontal} Lagrangian subspaces
$\y=constant$, each is identified with an element of the group of
phase space translations, which includes the identity, since
the identity subspace is defined as $\Vxi=0$. On the other hand,
the {\it vertical} subspace, $\x=\vct 0$, defines the canonical
reflection through the origin, $\x^- \mapsto \x^+=-\x^-$ (or
inversion), since all the chords for this transformation are
centred on the origin (see \cite{AlmBro06} or \cite{emaranha} for
further discussion.)

We can now, in analogy to (\ref{daction1}), interpret the centre
action $S(\x)$ in the semiclassical Wigner function
(\ref{semiwigner}) as 
\begin{equation} 
S(\x)= \int^\x \y(\x')\cdot \rm d \x'=
\int^\x \Vxi(\x')\wedge \rm d\x'. 
\label{doubleaction} 
\end{equation}
The integral is evaluated along a path on the Lagrangian submanifold
$\Lambda_\psi$ in double phase space,  
from some point on its intersection with the $\x$-plane.
(This intersection reproduces the single torus $\mathcal{L}_\psi$.)
The integral is independent of the path on
$\Lambda_\psi$, because $\Lambda_\psi$ is Lagrangian. We thus
obtain the chord (\ref{chorderiv}) by taking the derivative of
(\ref{doubleaction}).

The chord function is the Fourier transform of $W(\x)$. If this
transform of the semiclassical Wigner function is performed
within the stationary phase approximation, the semiclassical
expression for the chord function has a phase,
$\sigma(\Vxi)/\hbar$, such that the chord action, $\sigma(\Vxi)$,  
is the Legendre transform of the centre action, $S(\x)$.
It can be defined directly in terms of a similar integral to
(\ref{doubleaction}), with the roles of $\x$ and $\Vxi$ reversed:
\begin{equation} 
\sigma(\Vxi)=\int_0^{\Vxi} \x(\Vxi')\wedge \rm d \Vxi'=
-\int_0^{\J\Vxi} \x(\y')\cdot \rm d \y'= \sigma'(\y).
\label{doubleaction2} 
\end{equation} 
The action $\sigma(\Vxi)$ is, of course,
the same as appeared in the semiclassical theory for the chord
function (\ref{semichord}). When this theory is transported into
double phase space, it is often simpler to deal with
$\sigma'(\y)$. Then, within this formalism, the semiclassical
expression, for (each branch of) the pure state Wigner function or
chord function, assumes a generalized WKB form, derived by
Van Vleck \cite{Van Vleck}.

So as to treat the unitary evolution of the density operator,
which preserves the purity of
the state, $| \psi \rangle \langle  \psi |$, 
we need to consider the corresponding
classical evolution of both the tips of each chord, $\x^-$ and
$\x^+$, lying on a $2N$-dimensional lagrangian torus. Taking
account of the sign change of $p^-$, in the definition of double
phase space, we find that the double phase space Hamiltonian must
be \cite{OsKon} 
\begin{equation} I\!\!H_U(\X)=H(\x^+)-H(\x^-)=H(\x-\J
\y/2)-H(\x+\J \y/2). 
\label{Heisham} 
\end{equation} 
This Hamiltonian dynamics evolves lagrangian submanifolds in
double phase space, which correspond to pure density operators 
satisfying the Liouville-Von Neumann equation. The
explicit formulae for the semiclassical evolution of the Wigner
function are given in \cite{RiosOA, OsKon}, whereas the evolving
action (\ref{HJ_chord}) of the chord function is presented in
\cite{AlmBro06}. Reinterpreted as an evolving action in double
phase space, (\ref{HJ_chord}) assumes the form of an ordinary
Hamilton-Jacobi equation for $\sigma'(\y)$:
\begin{equation}
-\frac{\partial \sigma}{\partial t}(\Vxi,t) = -\frac{\partial
\sigma'}{\partial t}(\y,t) = I\!\!H(\frac{\partial
\sigma'}{\partial\y}, \y) = I\!\!H(\J \frac{\partial
\sigma}{\partial\Vxi}, \J\Vxi). \label{HJ_dchord}
\end{equation}
The difficulty lies in the neighbourhood of caustics of the initial state, 
which require more sophisticated semiclassical treatment.

According to the discussion in section 3, the evolution of the
amplitudes, in the decomposition of either the Wigner function or
the chord function, relies on the previous specification 
of translated action variables, for the lagrangian manifold, 
$\mathcal{L}$, corresponding to a semiclassical state. The
corresponding lagrangian submanifold is now a product, 
$\Lambda=\mathcal{L}^- \times \mathcal{L}^+$. Thus, the quantized
double torus is defined as the intersection of all the level
submanifolds of the $2N$ variables $\mathcal{I}_n^\pm$, defined by
(\ref{transaction1}), or 
\begin{equation}
\mathcal{I}_n^\pm(\X)=\mathcal{I}_n(\x\mp \J\y).
\label{transaction2} 
\end{equation} 
The fact that we are dealing with
projection operators restricts the Bohr-level for each pair of
variables $(\mathcal{I}_n^+(\X), \mathcal{I}_n^-(\X))$ to be the
same. 
\footnote{If this condition is relaxed, the present
semiclassical theory is immediately extended to include the
propagation of dyadic operators, corresponding to 
pairs of different eigenstates of $N$ commuting operators.}

Let us then consider the family of actions, $S(\x,
\mathcal{I}^\pm)$ or $\sigma(\Vxi, \mathcal{I}^\pm)$, evolving
classically for all possible constant values of the action
variables, $\mathcal{I}^\pm$. This is known as the complete
solution of the Hamilton-Jacobi equation \cite{Arnold}. Then, a
simple extension to double phase space of the usual canonical
formalism implies that \begin{equation} \fl |\det \frac{\partial^2\sigma'(\y,
\mathcal{I}^\pm, t)}{\partial\y \partial\mathcal{I}^\pm}|=
\big|\det \frac{\partial\mathcal{I}^\pm}{\partial\x}\big|^{-1}
\;\;\;{\rm and}\;\;\;\; |\det \frac{\partial^2S(\x,
\mathcal{I}^\pm, t)}{\partial\x
\partial\mathcal{I}^\pm}|= \big|\det
\frac{\partial\mathcal{I}^\pm}{\partial\y}\big|^{-1}.
\label{actder} \end{equation} Combining (\ref{actder}) with the expressions
(\ref{amplitude}) and (\ref{ampl2}) for the semiclassical
amplitudes, leads to \begin{equation} \fl a(\x,t)= |\det \frac{\partial^2S(\x,
\mathcal{I}^\pm, t)}{\partial\x \partial\mathcal{I}^\pm}|^{1/2}
=|\det \frac{\partial^2\sigma'(\y, \mathcal{I}^\pm, t)}{\partial\y
\partial\mathcal{I}^\pm}|^{1/2}
 =\alpha'(\y,t)=\alpha(\Vxi,t),
\label{ampl3} \end{equation} where this equality (\ref{ampl3}) between the
centre and the chord amplitudes holds only at a specific double
phase space point $\X=(\x, \y)$ on the double torus $\Lambda$, as
pointed out in the discussion at the end of section 2. Therefore,
the amplitudes of the evolving Wigner and chord functions are
entirely determined by the complete solution of the respective
Hamilton-Jacobi equations, in full analogy to the ordinary
evolution of WKB semiclassical states in the Schr\"odinger
formalism, derived by Van Vleck \cite{Van Vleck}.

It should be pointed out that our use of double phase 
concerns only the semiclassical approximation to the evolution 
generated by the Liouville-Von Neumann equation and susequently the
full Lindblad equation. At each instant, either the Wigner function, 
or the chord function are defined in the standard way, 
as the traces \eref{covW}, or \eref{chordfunction}, 
in terms of the single phase space of centres,
$\x$, or chords $\Vxi$, respectively. Though it may be tempting to
define an enlarged quantum evolution for superoperators
in direct correspondence with double phase space, no such 
generalization is treated here.  

Now, we finally turn to the semiclassical theory for markovian
evolution, as discussed in the previous section. We immediately
recognize in (\ref{decfl}) the same structure as that of the
double phase space Hamiltonian (\ref{Heisham}), that is:
\begin{equation}
I\!\!L(\X)=L(\x^+)-L(\x^-)=L(\x-\J \y/2)-L(\x+\J \y/2).
\end{equation}
This double phase space Lindblad function is the basic ingredient
in the decoherence functional, which is now defined along a single
trajectory in double phase space, generated by $I\!\!H_U(\X)$: 
\begin{equation}
D\{\X(t)\} :=  \sum_k\;\int_0^t dt'\;|I\!\!L_k(\X(t'))|^2  \ .
\label{decfl2} 
\end{equation} 
The square root of this functional can now be
interpreted as a time dependent {\it length} of the double phase
space vector, $\X(t)$, with $D\{(\x, \y\!=\!0)\}=0$ for all time, instead
of a {\it distance} between a pair of of single phase space
points.

In conclusion, we can interpret the conservative semiclassical
evolution of the chord function (\ref{chit}) entirely within the
double phase space picture. Indeed, this has assumed the same form
as general Van Vleck evolution, with the centre variables $\x$
playing the role of positions, while $\y=\J\Vxi$ stand for the
momenta. Following this analogy, the Wigner function substitutes
the Schr\"odinger wave function and the chord function is its
Fourier transform.

The only new element that has been added is the action of the
decoherence functional: The amplitude of the chord
function away from the origin progressively decays in time. This
quenching of the long chords can only be partially incorporated in
the semiclassical Wigner function, obtained from the alternative
projection of the same Lagrangian submanifold, $\Lambda(t)$, in
the limit of very short times. Indeed, it is only through this
submanifold that we can ascribe specific chords to each centre
$\x$. For finite times, it is better to calculate the Wigner
function as a full convolution, according to the discussion in the
previous section.

So far, that is, in the semiclassical theory for quantum unitary
evolution, the concept of double phase space may be considered to be 
somewhat redundant, because everything can be described in terms of pairs
of hamiltonian trajectories in single phase space. However, and
this is the fundamental point, for dissipative markovian systems,
we can identify the dissipative term in the full semiclassical
master equation (\ref{P9}), that is,
\begin{equation}
\fl I\!\!H_L(\X) :=- \sum_k \;\mathcal{I}m \; L_k(\x_j^+)L_k(\x_j^-)^*
=\sum_k\; \mathcal{I}m \; L_k(\x-\J\y/2)L_k(\x+\J\y/2)^* \ , 
\label{DL}
\end{equation}
as a new term of the double phase space Hamiltonian.
Indeed, the introduction of (\ref{SCcomm}), together with the
generalized version of (\ref{P9}) for several Lindblad operators,
into the chord representation of the master equation
(\ref{Lindblad}) results in the semiclassical evolution equation,
\begin{eqnarray} 
\label{partialchi}
\frac{\partial \chi_j}{\partial t} (\Vxi,t) = && \Big\{-{i\over
\hbar} \Big[H(\x_j^+)-H(\x_j^-)
-\sum_k \mathcal{I}m \Big(L_k(\x_j^+)L_k(\x_j^-)^*\Big) \Big]\nonumber\\
&& -{1\over{2\hbar}} \sum_k\Big|L_k(\x_j^+)-
L_k(\x_j^-)\Big|^2\Big\}\; \chi_j(\Vxi,t),
\end{eqnarray}
for each branch of the chord function, recalling that
$\x^\pm\equiv\x^\pm(\Vxi)=\x\pm\Vxi =\x\mp\J\y$.

In this way, for open dissipative systems, we can consider the
total Hamiltonian function on double phase space
$\bf{R}^{2N}\times \bf{R}^{2N}$ as given by
\begin{equation}
\label{markham} I\!\!H(X) = I\!\!H_U(X) + I\!\!H_L(X) \ ,
\end{equation}
where $I\!\!H_U(\X)$ and $I\!\!H_L(\X)$ are given respectively by
equations (\ref{Heisham}) and (\ref{DL}), for functions $H(\x)$
and $L_k(\x)$ on simple phase space $\bf{R}^{2N}$.
The particular combination of Lindblad functions, 
which we have recognized in \eref{DL} as a new term in the double Hamiltonian, 
also appears as the integrand of the {\it phase functional}
in Strunz's path inegral \cite{Strunz}, but there,
in the absense of double phase space, it lacks an interpretation.
Indeed, it is the very fact that $I\!\!H_L(\X)$ cannot be related to a
Hamiltonian in simple phase space in the same way as
$I\!\!H_U(\X)$, which now establishes the double phase space formalism 
as wholly indispensable. Therefore, for open dissipative systems, a
trajectory of the full Hamiltonian $I\!\!H(\X)$ in double phase
space is not equivalent to a pair of trajectories of a Hamiltonian
in simple phase space, as in the semiclassical theory for closed
systems.

\section{Dissipative semiclassical evolution}

To simplify our study, we here restrict the Lindblad operators to
be linear functions of positions and momenta, that is, given by
(\ref{lin}). The dissipative term (\ref{DL}) in the double phase
space Hamiltonian is then rewritten as 
\begin{equation} 
I\!\!H_L(X)= -\gamma\ \x
\cdot \y = -\gamma\ \Vxi\wedge \x , \label{DLL} \end{equation} defining the
dissipation coefficient, 
\begin{equation} 
\gamma= \sum_k \;
\Vl'_k\wedge\Vl''_k. 
\end{equation} 
The contribution of this term to
Hamilton's equations in double phase space is 
\begin{equation} \dot{\x} =
-\gamma \ \x \ \ \ , \ \ \dot{\y}=\gamma \ \y \ \ \ , \ \ \
(\dot{\Vxi} = \gamma \ \Vxi), \label{clasdis} 
\end{equation} 
so that the dissipative evolution in double phase space 
is always hyperbolic: contraction on the lagrangian subspace
$\bf{R}^{2N}\equiv\{\x\}$ (the identity plane, i.e. the centre
phase space) and expansion on its conjugate
$\bf{R}^{2N}\equiv\{\y\}$ (and hence of the chord space), or vice
versa. The fact that these are precisely the subspaces which
support the Weyl and the chord representation singles them out as
privileged choices for the description of quantum markovian
processes.

From here on, we unleash the centre space from its
interpretation as single phase space. This aspect has been
previously addressed in the review of Balazs and Jennings
\cite{BalazsJen}, but now it becomes an intrinsic feature of the
present theory. For, it is by recognizing, instead, the centre
space as a lagrangian subspace of the double phase space, that we
free symplectic areas and volumes on the centre space from being
conserved. In this way, dissipation finds a place in a strictly
real Hamiltonian theory. \footnote{However, the symplectic
structure on the centre space remains indispensable, as concerns 
its role in the definition of the Weyl symbols. 
The evolution of this representation follows from the motion
of the density operator, not from an evolution of
the symplectic structure on the centre space, which is not
preserved by the hamiltonian flow of (\ref{Heisham}), even if
$L\equiv0$, as long as $H$ is not quadratic.}

For the case where $\hat{L}$ is the annihilation operator
$\hat{L}=\hat{a}$, then $\Vl'=(0,2^{-1/2})$ and
$\Vl''=(2^{-1/2},0)$, so $\gamma >0$ and the centre motion is
contractive, while the spacing between neighboring chords expands
with time. In the case where $\hat{L}$ is the creation operator
$\hat{a}^{\dag}$, the opposite happens. The optical master
equation (\ref{opt}) combines both the creation and annihilation
operators, but in such a way that $\gamma
>0$, so the centre motion is contractive (strictly dissipative).
In the example of a two level atom coupled to a bath of photons in
a single field mode, this indicates that, although stimulated
emission compensates absorption, spontaneous emission leads to an
irreversible loss.

The double phase space formulation maintains the correspondence of
the evolving density operator, $\oprho(t)$, to a time-dependent
lagrangian submanifold, $\Lambda(t)$, even in the presence of
dissipation. It is true that, unlike $I\!\!H_U(\X)$, the new
dissipative term, $I\!\!H_L(\X)$, of the double phase space
hamiltonian destroys the factorization of the initial double
lagrangian submanifold into single phase spaces. However, the
generalized Van Vleck form of the evolution, which was shown to
hold for phases and amplitudes in the previous section, is in no
way restricted to product tori. Thus, there is no obstacle to the
immediate generalization of the present theory.

The evolving lagrangian submanifold, $\Lambda(t)$, can in
principle be described by an action function that measures its
symplectic area with respect to any (double phase space)
lagrangian coordinate subspace, such as the position subspace,
$\vct Q=(q^-, q^+)$, even though we can no longer describe
$\Lambda(t)$ locally as $p^\pm=\pm\; \partial s(q, t)/\partial
q^\pm$, with the same action function, $s(q)$, for $q^-$ and
$q^+$. Nonetheless, the Weyl and the chord representations, in
terms of the double phase space variables, $\x$ and $\Vxi=-\J \y$
respectively, have privileged roles: The former continues to be
interpreted as a quasiprobability, whereas the evolution of the
latter exhibits decoherence and diffusion in a specially simple
form, by merely quenching the amplitude of the various branches of
the chord action function (\ref{doubleaction2}). 

The evolution of
each branch is then obtained from the same Hamilton-Jacobi
equation (\ref{HJ_dchord}) as before, except that now we replace
the closed double Hamiltonian (\ref{Heisham}) by the full
markovian double Hamiltonian (\ref{markham}). Spelled out in terms
of the single space functions, for $I\!\!H_L(X)$ given by
(\ref{DLL}), this becomes
\begin{equation}
-\frac{\partial \sigma}{\partial t}(\Vxi,t) =
H\big(\J\frac{\partial \sigma}{\partial \Vxi} +
\frac{\Vxi}{2}\big) - H\big(\J\frac{\partial \sigma}{\partial\Vxi}
- \frac{\Vxi}{2}\big) + \gamma \Vxi \cdot \frac{\partial
\sigma}{\partial \Vxi}  \;\; . \label{HJ-disschord}
\end{equation}
Recalling the simple form of the action
(\ref{doubleaction2}) in double phase space, the evolution of the
lagrangian submanifold, $\Lambda(t)$, is just given by $\x(\y, t)=
-\partial \sigma'(\y, t)/\partial\y$. 

The semiclassical approximation for the evolving chord function
is still given by (\ref{chit}) for each of the
branches of the chord function.
The action function  evolves according to (\ref{HJ-disschord}), 
whereas the amplitude of the decoherenceless factor, 
${\chi^0}_j(\Vxi,t)$ is specified by (\ref{ampl3})  . 
In the general case, where the
Hamiltonian is not quadratic, the closed evolution of the
decoherenceless factor of the chord function in (\ref{chit}) is
not obtained from the single phase space trajectories,
$\Vxi(-t)$, generated by the classical Hamiltonian, $H(\x)$, i.e.
${\chi_j}^0(\Vxi,t)\neq{\chi_j}^0(\Vxi(-t), 0)$.

Having reinterpreted the decoherence functional (\ref{decfl})  in
double phase space as (\ref{decfl2}), we now obtain the full
semiclassical markovian evolution for each branch of the chord
function, including dissipation, in the same form (\ref{chit}) as
before, even though the evolving ingredients can no longer be
interpretated in single phase space. Because of the linearity
assumed for all the Lindblad operators (\ref{lin}), the
decoherence functional (\ref{decfl2}) or (\ref{decfl}) takes the
explicit form,
\begin{equation}
D\{\x_j^+(t),\x_j^-(t)\}= \sum_k \int_0^t dt'
\Big[|\Vl'_k\cdot\overline{\Vxi}_j(\Vxi,t')|^2 +
|\Vl''_k\cdot\overline{\Vxi}_j(\Vxi,t')|^2\Big], \label{dissdec}
\end{equation}
for each branch of the chord function. It is the classical
evolution of the full double phase space vector,
$\X_j=(\x_j^-(\Vxi), \;\x_j^+(\Vxi))$ (or, in alternative
coordinates, $\X_j=(\x_j(\Vxi), \;\y=\J\Vxi)$) that determines the
decoherence functional, i.e. $\overline{\Vxi}_j(\Vxi, -t')$ is
obtained by multiplying the $\y$-component of $\X_j(t)$ by $-\J$.

It is only the further restriction to a quadratic Hamiltonian that
forces all the chord projections in double phase space to evolve
in the same way, independently of each centre, $\x_j(\Vxi)$. Then
the evolution of all the semiclassical branches can again be
united into (\ref{linn}), with the only difference that now
$\Vxi(t)$ is obtained from the expansive linear equation
\cite{BroAlm04}: 
\begin{equation} 
\dot{\Vxi}=\J\frac{\partial
H}{\partial\Vxi}+\gamma\; \Vxi. \label{diss'} 
\end{equation} 
This is just one
of Hamilton's equations for the full double phase space
Hamiltonian (\ref{markham}). Since (\ref{linn}) coincides with the
exact solution of the Lindblad equation for quadratic $H(\x)$ and
linear $L(\x)$, it follows that the present semiclassical theory
is exact in this limit, even in the presence of dissipation.

There is a subtle distinction to be noted in the derivation of the
semiclassical evolution of a dissipative Markovian system, with
respect to the theory in section 5. The approximation there could
be interpreted as the short time limit of a conceivable periodic
system, wherein the internal Hamiltonian and the Lindblad
operators were alternatively turned on and off. The full
dissipative approximation is now derived in the same way, if we
include the new dissipative part in the double Hamiltonian.
Even though the Trotter ansatz \cite{Schulman} can still be invoked,
this mathematical procedure is now devoid of its
interpretation as a conceivable periodic system, because
we cannot physically switch off the decohering part of the Lindblad equation
without eliminating the dissipative part of the Hamitonian: 
In the present context, it is
possible to have decoherence without dissipation, but not the
other way around. In any case, the extension of the semiclassical
approximation to dissipative systems is also exact in the quadratic limit.

The evolution of the semiclassical Wigner function now follows
through the derivation in section 5: The Fourier transform of
(\ref{chit}) will be a convolution of ${W_j}^0(\x,t)$, the Fourier
transform of ${\chi_j}^0(\Vxi,t)$, with a widening window which
coarse-grains over the interferences of the Wigner function.
Beyond the decoherence time, the Fourier transform of the chord
function, i.e. the Wigner function becomes smooth and
classical-like. From then on, the classical motion 
on this centre space is given by Hamilton's (single phase space) equations, 
with the addition of the purely dissipative term (\ref{clasdis}): 
\begin{equation}
\dot\x=\J\frac{\partial H}{\partial\x}-\gamma \x
\label{diss''}
\end{equation} 
(to which diffusion is always added by convolution).
The reason is that this is an invariant subspace for unitary
double phase evolution \cite{AlmBro06}, a property which is not
altered by dissipation. Since the decoherence functional has, at
this stage, effectively cancelled all large chords, the physical
interest is concentrated on this plane. In this regime the
evolution of the decoherenceless factor of the Wigner function in
(\ref{wigt}), can be pictured as purely classical,
${W^0}_{j}(\x,t)={W^0}_{j}(\x(-t), 0)$, with $\x(-t)$ obtained from
(\ref{diss''}).

It might seem strange that the semiclassical solution of the
dissipative master equation for the chord function becomes exact
in the case of linear Lindblad phase space functions and a
quadratic Hamiltonian. The latter condition is familiar on its
own, but the exact Lindblad term (\ref{exactdiss}) is not in the
same form as the semiclassical approximation (\ref{semilin}), so
that it is not evident that the same limiting behaviour is
obtained. However, let us, in this case, reinterpret the discrepant term as part
of a quantized Hamiltonian superoperator for a Schr\"odinger-like
equation corresponding to double phase space.
Then $\widehat{\x}$ and $\widehat{\y}$ will be operators corresponding respectively 
to positions and momenta in double phase space, 
such that $\widehat{\x}=i\hbar\frac{\partial }{\partial\y}$.
It follows that $\chi(-\J\y)$ can then be interpreted 
as a {\it double wave function}, so that 
\begin{equation} \fl
(\Vl'\wedge\Vl'') \; \Vxi\cdot\frac{\partial \chi}{\partial \Vxi}
= (\Vl'\wedge\Vl'') \; \y\cdot\frac{\partial }{\partial
\y}\;\;\chi(-\J\y) = -{i\over\hbar}(\Vl'\wedge\Vl'')
\;\widehat{\y}\cdot\widehat{\x}\;\;\chi(-\J\y), 
\end{equation} 
and the action
of this Hamiltonian superoperator on a semiclassical branch of the
chord function, to first order in $\hbar$, 
\begin{equation}
-{i\over\hbar}(\Vl'\wedge\Vl'')
\;\widehat{\y}\cdot\widehat{\x}\;\;\chi_j(-\J\y)= -(\Vl'\wedge
\Vl'')\;\frac{\partial\sigma_j}{\partial \Vxi}\cdot \Vxi\;\;
\chi_j(\Vxi), 
\end{equation} 
is the same as in (\ref{semilin}).

\section{Semiclassical markovian propagators}

The present semiclassical picture for markovian
evolution of the density operator has dealt directly with the
chord function and its Fourier transform, the Wigner function. The
problem that must now be addressed is that, whereas the
decoherence functional quickly quenches the contribution of large
chords, the remaining {\it classical} small chord region lies in
the neighborhood of a caustic for both the chord function and the
Wigner function. In other words, the region of the lagrangian
submanifold, $\Lambda(t)$, lying close to the centre subspace
$\y=0$ in double phase space, projects singularly onto both the
$\x$-subspace and the $\y$-subspace. Hence, the direct
semiclassical theory above is only applicable to the initial
stages of markovian evolution, before the long chords are mostly
quenched. 
\footnote{On the other hand, it is the essentially
nontransversal intersection of the double torus,
$\Lambda=\mathcal{L}(\x^-) \times \mathcal{L}(\x^+)$, 
with the central subspace, along $\mathcal{L}(\x)$, that allows for a
nontrivial Wigner function even after decoherence, such as is
observed in the quadratic limit. (The single torus, $\mathcal{L}(\x)$,
has the same form as the factor tori in the respective spaces, $\x^\pm$.)} 
Though the further processes of
dissipation and diffusion proceed continuously through the initial
stage and beyond, their direct description requires an advanced
semiclassical treatment of markovian evolution in the caustic
region, beyond the scope of the present theory.

One way out of this problem is to consider alternative lagrangian
submanifolds, which do not have caustics, at least initially. This
is the approach adopted in \cite{AlmBro06}. Instead of the
submanifold $\Lambda(t)$, which corresponds to $\oprho(t)$, but
has undesirable caustics, we evolve the submanifolds
$\x=constant$, or $\y=constant$, corresponding respectively to
unitary reflection, or translation operators, $\opR_\x$, or $\opT_{\Vxi}$. 
The $\x=constant$ submanifold has no caustic in its $\y$-projection,
while the $\y=constant$ submanifold is free from caustics in its
$\x$-projection. Furthermore, a finite time must pass before the
evolution generated by the double Hamiltonian can bend either of
these submanifolds sufficiently to produce caustics, until which
time both the chord representation of $\opR_\x(t)$ and the centre
representation of $\opT_{\Vxi}(t)$ will be represented in the
simple semiclassical forms (\ref{semiwigner}) and
(\ref{semichord}). In other words, $\tilde{R}_\x(\Vxi, t)$, the
chord representation of the reflection operator, and
$T_{\Vxi}(\x,t)$, the centre representation of the translation
operator, respectively, will both have a single semiclassical
branch.

In this way we obtain a semiclassical approximation to the
evolution, whether unitary or markovian, by inserting the
approximate evolved operators in the exact relations
\cite{AlmBro06}, 
\begin{equation} W(\x,t)= \int \frac{\rmd\Vxi}{(2\pi\hbar)^N}
\>\chi(\Vxi) \> T_{\Vxi}(\x,t), \label{chord-centre} 
\end{equation} 
or 
\begin{equation}
\chi(\Vxi,t)=\int  \frac{\rmd \x}{(2\pi\hbar)^N} \> W(\x)\>
2^N\;\tilde{R}_{\x}(\Vxi,t).
\label{centre-chord} 
\end{equation} 
Unlike these
{\it mixed} propagators, which involve both centres and chords,
the semiclassical expressions for {\it direct} propagators of
Wigner functions \cite{Dittrich} necessarily involve uniform
approximations through caustics, even in the simple case of
unitary evolution.

In the present context, it is (\ref{centre-chord}) that should be
chosen, because the markovian evolution of a reflection operator,
initially represented by 
\begin{equation} 
2^N\;\tilde{R}_{\x}(\Vxi)= \exp
{\left({i\over\hbar} \x \wedge \Vxi\right)}, 
\label{planewave} 
\end{equation}
is approximated semiclassically within the chord representation in
the same way as a single branch of the chord function,
$\chi(\Vxi)$ in (\ref{semichord}). This holds for arbitrary
Hamiltonians, together with linear Lindblad operators, 
self-adjoint or not. Thus, the approximate evolution of (\ref{planewave})
has precisely the same form as (\ref{chit}), i.e.
\begin{equation}
\label{Rxt} 
\tilde{R}_\x(\Vxi,t)= {\tilde{R}_\x}^0(\Vxi,t)\;\;
\exp{\Big\{-\frac{1}{2\hbar}\sum_k
\int_0^tdt'\big|\Vl_k\cdot\overline{\Vxi}(-t')\big|^2\Big\}},
\end{equation}
where $\overline{\Vxi}(t)$ is the trajectory of the initial chord,
$\overline{\Vxi}(0)=\Vxi$, obtained from the $\y$ component of the
evolving double phase space vector, $\X=(\x,\; \y=\J\Vxi)$, and the
Lindblad coefficients (\ref{lin}) are in general complex. Here,
the decoherenceless factor is 
\begin{equation} 
{\tilde{R}_\x}^0(\Vxi,t)=
{\tilde{R}_\x}^0(-\J\y,t)=
\frac{1}{2^N}\Big|\frac{\partial^2\sigma'_\x (\y, t)}{\partial\y
\partial\x}\Big|^{1/2} \exp {\left({i\over \hbar}\sigma'_\x (\y, t)
\right)}. \label{declR} 
\end{equation}

The evolution of ${\tilde{R}_\x}^0(\Vxi,t)$ does not coincide
with the unitary evolution presented in \cite{AlmBro06}, because of 
the dissipative term in the Hamiltonian.
In short, we merely substitute the transported action variables
$\mathcal{I}^\pm$, that previously defined the evolving lagrangian surface 
in double phase space, by $\x$ in (\ref{ampl3}) and the chord action,
$\sigma_\x (\Vxi, t)=\sigma'_\x (\y, t)$ is governed by the
Hamilton-Jacobi equation (\ref{HJ-disschord}). Evidently, the
initial action is $\sigma'_\x (\y, 0)=\x\cdot\y$, so that the
initial semiclassical expression coincides with (\ref{planewave}).
It should be noted that, at the chord origin, 
${\tilde{R}_\x}^0(\Vxi,t)=2^{-N}$ for all times,
and multiplication by the exponential of the decoherence
functional does not alter this value. Therefore, 
normalization is preserved, according to (\ref{normalization}). 

The conditions for the derivation of (\ref{Rxt}) must now be
analyzed. Recall that the exponential of the decoherence
functional was assumed to be a smooth (non-oscillatory) function
in our previous derivation of the semiclassical chord function
(\ref{chit}). For small chords, which are now in focus, this
smoothness assumption is falsified as the decoherence time is
reached, i.e. the time for the volume of the decoherence factor to
shrink to that of a coherent state. This establishes the duration
beyond which this chord propagator is valid. Nonetheless,
the evolved propagator can be reexpressed in terms of (static)
reflection operators, i.e. in the Weyl representation
and this whole procedure can then be iterated indefinitely.

It is precisely in this region, where the direct semiclassical
approximation for the chord function (\ref{chit}) is singular,
that the much simpler form, \begin{equation} 2^N\;{\tilde{R}_\x}^0(\Vxi,t)=\exp
{\left({i\over \hbar}\x (t)\wedge \Vxi \right)} , \end{equation} can be
employed \cite{AlmBro06}. Here, $\x(t)$ is the trajectory, issuing
from $\x(0)=\x$, that integrates (\ref{diss''}) in single phase
space, because of the restriction to the invariant subspace
$\y=0$. This approximation implies that the evolved reflection
operator is still represented by a vertical subspace in double
phase space. This is indeed true for motion generated by a
quadratic Hamiltonian and even a general closed double
Hamiltonian, $I\!\!H_U(X)$, of the form (\ref{Heisham}), leads to
a submanifold whose tangent space is vertical at the identity
subspace, throughout the motion, as discussed in \cite{AlmBro06}.
Fortunately, the new open dissipative term, $I\!\!H_L(X)$, of the
double Hamiltonian (\ref{DLL}) preserves this feature. 
In essence, the reason for this is that the evolution can be linearized
in the neighbourhood of a generalized reflection 
even in the presence of dissipation. 

To evaluate the corresponding approximation for the decoherence
functional, we expand the double Hamiltonian (\ref{markham}) as 
\begin{equation}
I\!\!H(\x,\y) = \frac{\partial H}{\partial \x}(\x)\wedge \y -
\gamma \; \x \cdot \y, 
\end{equation} 
which holds to second order in $\y$. Then,
Hamilton's equation for the $\y$-motion in the neighborhood of the
invariant centre subspace is just 
\begin{equation} 
\dot\y(t) \ = \
[2\mH_2(\x(t))\J+\gamma]\; \y, \label{ZZ} 
\end{equation} 
where the independent centre motion $\x(t)$ is determined by the 
non-hamiltonian classical equation \eref{diss''} and $\mH_2(\x)$ is the
Hessian matrix for $H(\x(t))$. The quadratic form
$\y\cdot\mH_2(\x(t))\y$ can be interpreted as a local Hamiltonian for
the motion transverse to the centre subspace, though it receives
an extra boost from the dissipation coefficient, $\gamma$. Given
an initial point $\X=(\x, \;\y\!=\!\J\Vxi)$, whose orbit is
assumed to remain close to the identity subspace within the time $
t $, we thus obtain its chord evolution as 
\begin{equation} 
\Vxi_t(\x,
\Vxi)=\G_t(\x) \;\Vxi, 
\end{equation} 
where the classical propagation matrix
is
\begin{equation} 
\G_t(\x):=\mathop{\lim}_{N\rightarrow\infty} \; 
\prod_n^N\; \exp\Big[\frac{t}{N}[2\mH_2(\x(\frac{nt}{N}))\J+\gamma] \Big],
\end{equation} 
which, for small times is approximately
\begin{equation} 
\G_t(\x)\approx \ \exp \Big[\int_0^t {\rm d}t' \;
(2\mH_2(\x(t'))\J+\gamma)\Big]. 
\end{equation} 
The only difference between
this approximation and the exact evolution for the quadratic case
in \cite{BroAlm04} is the $\x$-dependence for $\mH_2$ and, hence,
the need for an integral in the definition of $\G_t$.
Furthermore, the present definition incorporates the dissipation
coefficient, $\gamma$, from the full double Hamiltonian
(\ref{markham}).

We are dealing here with trajectories that remain very close to
the invariant centre subspace, where ${D\{\x(t),\; \y=0\}}=0$. In
other words, within the short chord approximation, the pair of
trajectories, which define the decoherence functional, become
indistinguishable from a single classical trajectory, so that 
\begin{equation}
D\{\x(t), \y\rightarrow 0\}= \; {\Vxi \cdot \M_t(\x) \Vxi \over 2}, 
\end{equation} 
a simple quadratic form in the variables transverse to this centre
subspace, specified by the evolving matrix, 
\begin{equation} 
\M_t(\x)= \sum_k\int_0^t {\rm  d}t' \; \mathbf{G}^T_{t'}(\x)\vl_k \vl^T_k
\mathbf{G}_{t'}(\x). 
\end{equation} 
Here $(.)^T$ denotes the transpose of a
matrix, or a vector. This quenching exponent has exactly the same
form as the exact one for quadratic Hamiltonians \cite{BroAlm04}
along the centre subspace. The only difference lies in the
dependence of the classical chord propagator $\mathbf{G}_{t}(\x)$
on the local quadratic approximations of the Hamiltonian along the
trajectory $\x(t)$, which can be taken either in the forward direction,
starting at $\x$, or backwards from $\x(t)$. In conclusion, the
small chord approximation for the mixed propagator is 
\begin{equation}
2^N\;\tilde{R}_\x(\Vxi,t)=\exp {\left({i\over \hbar}\x (t)\wedge
\Vxi \right)} \;\exp\Big(-\frac{\Vxi \cdot \M_t(\x) \Vxi}{2\hbar}
\Big). \label{smchord} 
\end{equation}

This can now be inserted into (\ref{centre-chord}), to obtain 
\begin{equation}
\chi(\Vxi,t)=\int  \frac{\rmd \x}{(2\pi\hbar)^N} \> W(\x)\> \exp
{\left({i\over \hbar}\x (t)\wedge \Vxi \right)}
\exp\Big(-\frac{\Vxi \cdot \M_t(\x) \Vxi}{2\hbar} \Big).
\label{dechord} 
\end{equation} 
Here, we cannot use the semiclassical
approximation (\ref{semiwigner}) for the initial Wigner function,
$W(\x)$, in the above integral, because it is singular in the
small chord region, but improved uniform approximations
\cite{Ber77, Alm83} are valid in this range. A correct initial
normalization of the Wigner funtion guarantees the normalization
of the chord function for any subsequent time, according to
(\ref{normalization}). 

It is now worthwhile to review the full construction required to
follow the markovian evolution of a given initial pure state. By
inserting its Wigner function into (\ref{centre-chord}) along with
the semiclassically evolved propagator, specified by (\ref{Rxt})
and (\ref{declR}), we obtain the evolution of the chord function
for a time interval of the order of the decoherence time. The
amplitudes of all long chords will be strongly quenched in this
evolved chord function, so that its Fourier transform, the evolved
Wigner function, becomes smooth and positive. This can be further
evolved by inserting it into (\ref{centre-chord}) again. But now
there will be no large chord contribution, even right from the
start, because at $t=0$, equation (\ref{centre-chord}) reduces to
the inverse of the Fourier transform that we have just made.
Therefore, the much simpler evolution given by the small chord
approximation (\ref{dechord}) is now adequate. The markovian
evolution never {\it dequenches} the large chord amplitudes, so
that all further iterations of this procedure may safely rely on
the small chord approximation.

Let us now describe the evolutions that can be calculated within
the small chord approximation entirely within the Weyl
representation. In the limit of small propagation time,
$\M_t\rightarrow 0$, the Fourier transform of the evolving chord
function is exactly $W(\x', t)=W(\x(-t), 0)$, the classical
propagation of the Wigner function. This is the limit where the
Wigner-Wigner propagator becomes merely $\delta(\x'-\x(t))$.
Unless the Hamiltonian is quadratic, it is only valid because 
the long chords that have already been quenched out of the
propagator. For longer times, it is also possible to perform the
Fourier transformation of (\ref{dechord}) exactly, so as to obtain
the Wigner function as an evolving convolution with a gaussian
that broadens from an initial Dirac $\delta$-function. Following
\cite{BroAlm04}, this is given by 
\begin{equation} 
\fl W(\x',t)= \int
\frac{\rmd \x}{(2\pi\hbar)^N} \> {W(\x)\over\sqrt{\det
\M_t(\x)}}\; \exp\Big(-\frac{(\x- \x'(-t)) \cdot
\M'_t(\x)(\x-\x'(-t))}{2\hbar} \Big), \label{evwigner} 
\end{equation} 
in which $\x'(-t)$ is the backward trajectory of the evaluation
point, $\x'$. It is important to emphasize here, once again, that
this is a trajectory not of the single Hamiltonian $H(\x)$, but of
equation (\ref{diss''}), which adds the dissipative term,
$-\gamma\x$, to Hamilton's equation for the single Hamiltonian
$H(\x)$. The matrix 
\begin{equation} \
M'_t(\x):= -\J\M_t(\x)^{-1}\J 
\end{equation}
characterizes the Fourier transform of the exponential of the
decoherence functional in (\ref{dechord}). This is now a
broadening gaussian window, which coarse-grains the classical
evolution of the initial Wigner function. The same
coarse-graining, which accounted for the initial loss of quantum
coherence of an initial pure state, can now be interpreted as
resulting from a classical Langevin equation for Brownian motion
acting on a purely classical probability distribution in phase
space. This description is again the exact result in the quadratic case
\cite{BroAlm04} and it has been derived independently many times 
for the dampened harmonic oscillator \cite{Agarwal71,DodMan85,Sand87}. 
In the purely classical context, this picture for Brownian motion
can already be found in the review \cite{WangUhl45}.

It is tempting to extrapolate the approximate small chord
evolution (\ref{dechord}) beyond the decoherence time. After all,
the effect of this process is precisely to eliminate the
contribution of long chords, so that the passage to
(\ref{evwigner}) becomes more valid. Indeed, there would be no
obvious contradiction if the validity of (\ref{dechord}) were to
extend beyond the range allowed by our derivation, as is true in
the quadratic case. If we do insert (\ref{smchord}) into
(\ref{L7}), the integral to be evaluated becomes
\begin{eqnarray}
\fl (\hat{L}\opR_\x\hat{L}^{\dag} -
\frac{1}{2}\hat{L}^{\dag}\hat{L}\opR_\x -
\frac{1}{2}\opR_\x\hat{L}^{\dag}\hat{L})(\Vxi') = \nonumber\\
\fl \int\frac{d\Vxi''d\x''}{(2\pi\hbar)^{2N}} \exp {\left({i\over
\hbar}\x (t)\wedge \Vxi'' \right)} \;\exp\Big(-\frac{\Vxi'' \cdot
\M_t(\x) \Vxi''}{2\hbar}
\Big)\exp{\Big(\frac{i}{\hbar}[\x''\wedge(\Vxi'-\Vxi'')]\Big)}
\nonumber\\
\fl
 \Big\{\! L(\x''\! + {\Vxi'\over 2})L(\x''\! -{\Vxi''\over 2})^{*}
\!-\!\frac{1}{2}\big[L(\x''\! + {\Vxi''\over 2})L(\x''\! +
{\Vxi'\over 2})^{*} + L(\x''\! -{\Vxi'\over 2})L(\x''\! -
{\Vxi''\over 2})^{*}\big]\Big\}. \label{L8}
\end{eqnarray}
Recalling that the Lindblad functions $L(\x)$ are assumed linear,
we can now substitute $\x''$ by $i\hbar\J\partial/\partial\Vxi'$
within the brackets $\{\}$ in (\ref{L8}) and then integrate over
this variable to obtain $\delta(\Vxi'-\Vxi'')$. Thus, in this
case, the integral of (\ref{L8}) becomes
\begin{eqnarray}
\fl (\hat{L}\opR_\x\hat{L}^{\dag} -
\frac{1}{2}\hat{L}^{\dag}\hat{L}\opR_\x -
\frac{1}{2}\opR_\x\hat{L}^{\dag}\hat{L})(\Vxi') =\nonumber\\
\fl\left(i(\Vl'\wedge \Vl'') [\x(t)\wedge \Vxi'- \Vxi' \cdot
\M_t(\x) \Vxi']
-\frac{1}{2}[(\Vl'\cdot\Vxi')^2+(\Vl''\cdot\Vxi')^2]\right) \tilde{R}_{\x}(\Vxi',t)
\end{eqnarray}
if we neglect semiclassically small terms arising from the
derivatives of the Lindblad functions.

This is almost in the same form as (\ref{semilin}), except for the
term $i(\Vl'\wedge \Vl'') \; \Vxi' \cdot \M_t(\x) \Vxi'$, which is not
semiclassically small. Nonetheless, this term does disappear if the
dissipation coefficient $\gamma=(\Vl'\wedge \Vl'')=0$, 
so allowing us to carry through an
analogous derivation to that of the semiclassical chord function
in section 5. The result is that in this case the small chord
approximation (\ref{dechord}) is valid and the closed formula for
long term evolution of the Wigner function (\ref{evwigner}) can
be legitimately extrapolated far beyond the initial period for
decoherence. In the general dissipative case with nonquadratic
Hamiltonians, the qualitative picture of the Markovian evolution,
resulting from repeated iterations of (\ref{evwigner}) is the
same, but a single closed formula is not yet available.

\section{Discussions}

The point of view of this paper is that semiclassical Wigner functions 
or chord functions stand to double phase space as do semiclassical 
position and momentum wave functions to single phase space.
In strict analogy to the more familiar theory, each of these
conjugate representations is defined in its own subspace and
contains complete information concerning the quantum state, be it
pure or mixed. However, mixed systems demand a density operator
description, rather than as states in Hilbert space, which can be
provided by the Wigner function or the
chord function. For strictly unitary evolution, 
it is still possible to restrict consideration
to the single phase spaces on which these functions are originally
defined, but not for general quantum markovian processes.

No attempt has here been made to expand the framework 
of quantum mechanics itself. The Wigner function and the chord function 
are particular choices of representation 
for the evolving density operator, each in their own phase space.
It is only the identification of the various terms in these representations 
of the Lindblad equation with those of standard semiclassical 
wave function evolution, that leads to a generalized WKB-like solution 
for quantum markovian motion. 

The full double phase space Hamiltonian would have an unfamiliar
form if it were to be considered as the generator of motion for a
mechanical system in an ordinary multidimensional phase space,
but, once these peculiarities are understood, there results a
qualitative picture for generalized semiclassical evolution that
is pleasingly intuitive: The decoherence functional quickly
quenches the contribution of all large chords, just as in the
exact quadratic case \cite{BroAlm04}. Hence, after a short {\it
decoherence time}, we may restrict the analysis to the
neighbourhood of the centre subspace $\y=0$ in double phase space.
The fact that this is the subspace, where the Wigner function is supported, 
indicates that the conjugate pair of the
Weyl and the chord representations constitute a privileged frame
for the study of markovian evolution for quantum systems.

The various semiclassical representations of the density operator
are derived from a single Lagrangian submanifold in double phase
space, with its hamiltonian evolution. It so happens that the
product nature of this manifold implies that its intersection with
the identity subspace coincides with a caustic for both the Wigner
function and the chord function. We have shown in section 8 that a
simple adaptation of the present theory, obtained by choosing a set of
more favourable lagrangian submanifolds, furnishes an optimum
semiclassical propagator. This allows for appropriate small chord
approximations, leading back to a closed formula for the evolution
of the Wigner function (\ref{evwigner}). Thus, we again establish
contact with the exact markovian theory for quadratic
Hamiltonians.

Notwithstanding that the evolution of our propagator  
is derived in the semiclassical approximation, 
it will transport any kind of pure or mixed Wigner function. 
These may be initial semiclassical states,
but also (squeezed) coherent states, {\it Schr\"odinger cat
states}, or whatever. Though the analysis is harder in the
intermediate stage, between the initial loss of quantum coherence
and asymptotic classical motion, it can be conjectured that the
Wigner function becomes positive everywhere, even if it is doubtful
that the time for this will be independent of the initial state,
as in the quadratic case \cite{DioKief, BroAlm04}.

The fact that the present semiclassical theory is exact in the
case of a quadratic internal Hamiltonian, even in the presence of
linear Lindblad operators, can be considered as an indication that
it provides a useful generalization of this simple case. It will
certainly be necessary to make detailed comparisons of the
approximate semiclassical evolution to the direct integration of
the exact equation in the case of nonquadratic Hamiltonians.
However, the integration of the multidimensional partial
differential master equation is a considerable enterprise,
specially for the highly oscillatory Wigner functions described by
semiclassical theory. An alternative is to resort to quantum
monte carlo methods, such as in \cite{Percival}, but then the
comparison is merely between alternative approximations.

Our analysis has dealt only with Lindlad operators that are linear
functions of position and momentum operators. This may be
justified by picturing these operators as the quantum variables
that are responsible for the coupling to the environment, as in
the derivations of our standard example (\ref{opt}). Weak
coupling, as assumed for a markovian theory, often implies that a
linear approximation is valid, but this need not be so. Perhaps it
has been the very difficulty of obtaining a fully reliable general
picture of the evolution of the density operator of an open system
that has so far hampered the study of systems described by
nonlinear Lindblad operators. Further generalization of the
present theory to include this possibility is the subject of
ongoing work.

It is early to predict whether the insight provided within a full
semiclassical theory of non-unitary evolution will reveal features
which are beyond our present intuition. This can only be decided
by analyzing examples of increasing complexity within the present
framework, or by answering harder mathematical questions of the
formalism itself (e.g. as in \cite{Rios}), or perhaps both. 
For the present, we have gained
in understanding how dissipation and diffusion are emmerging
properties of the single continuous markovian evolution in double
phase space, which, in its initial stages, is best described as
decoherence. It is only when the contribution of long chords is
quenched onto the neighbourhood of the centre subspace that the
motion can meaningfully be described as dissipative.

\ack

Partial financial support from Millenium Institute of Quantum
Information, FAPERJ, PROSUL, CNPq, CAPES-COFECUB 
and UNESCO/IBSP Project 3-BR-06 is gratefully acknowledged.

\end{document}